\documentclass[web]{ieeecolor}
\usepackage{tmi}
\usepackage{cite}
\usepackage{amsmath,amssymb,amsfonts}
\usepackage{algorithmic}
\usepackage{graphicx}
\usepackage{textcomp}
\usepackage{epstopdf}
\usepackage{booktabs}
\newsavebox\CBox
\def\textBF#1{\sbox\CBox{#1}\resizebox{\wd\CBox}{\ht\CBox}{\textbf{#1}}}
\def\BibTeX{{\rm B\kern-.05em{\sc i\kern-.025em b}\kern-.08em
    T\kern-.1667em\lower.7ex\hbox{E}\kern-.125emX}}
\begin{document}
\clearpage

\twocolumn
\pagenumbering{arabic}
\setcounter{page}{1}
\title{Alleviating Class-wise Gradient Imbalance for Pulmonary Airway Segmentation}
\author{Hao Zheng, Yulei Qin, \IEEEmembership{Student Member, IEEE}, Yun Gu, \IEEEmembership{Member, IEEE}, Fangfang Xie, \\ Jie Yang, Jiayuan Sun, Guang-Zhong Yang, \IEEEmembership{Fellow, IEEE}
\thanks{This work was partly supported by 
	National Key R\&D Program of China (2019YFB1311503, 2017YFC0112700), Committee
	of Science and Technology, Shanghai, China (19510711200), Shanghai
	Sailing Program (20YF1420800), NSFC (61661010, 61977046, 62003208). 
	(Corresponding authors: Jie Yang, Jiayuan Sun and Guang-Zhong Yang.)}
\thanks{Hao Zheng, Yulei Qin, Yun Gu, Jie Yang and Guang-Zhong Yang are with the Institute of Medical Robotics, Shanghai Jiao Tong University, Shanghai, China. (e-mail: jieyang@sjtu.edu.cn; gzyang@sjtu.edu.cn)}
\thanks{Hao Zheng, Yulei Qin, Yun Gu and Jie Yang are with the Institute of Image Processing and Pattern Recognition, Shanghai Jiao Tong Univeristy, Shanghai, China.}
\thanks{Hao Zheng and Guang-Zhong Yang are with the School of Biomedical Engineering, Shanghai Jiao Tong Univeristy, Shanghai, China.}
\thanks{Fangfang Xie and Jiayuan Sun are with the Department of Respiratory and Critical Care Medicine, Department of Respiratory Endoscopy, Shanghai Chest Hospital, Shanghai Engineering Research Center of Respiratory Endoscopy, Shanghai, China. (e-mail: jysun1976@163.com)}
}

\maketitle

\begin{abstract}
Automated airway segmentation is a prerequisite for pre-operative diagnosis and intra-operative navigation for pulmonary intervention. Due to the small size and scattered spatial distribution of peripheral bronchi, this is hampered by a severe class imbalance between foreground and background regions, which makes it challenging for CNN-based methods to parse distal small airways. In this paper, we demonstrate that this problem is arisen by gradient erosion and dilation of the neighborhood voxels. During back-propagation, if the ratio of the foreground gradient to background gradient is small while the class imbalance is local, the foreground gradients can be eroded by their neighborhoods. This process cumulatively increases the noise information included in the gradient flow from top layers to the bottom ones, limiting the learning of small structures in CNNs. To alleviate this problem, we use group supervision and the corresponding WingsNet to provide complementary gradient flows to enhance the training of shallow layers. To further address the intra-class imbalance between large and small airways, we design a General Union loss function that obviates the impact of airway size by distance-based weights and adaptively tunes the gradient ratio based on the learning process. Extensive experiments on public datasets demonstrate that the proposed method can predict the airway structures with higher accuracy and better morphological completeness than the baselines. 
\end{abstract}


\section{Introduction}
Airway segmentation from computed tomography (CT) scans is used in a wide range of diagnostic and interventional procedures for lung diseases. As manual annotation is error prone, time-consuming, and requires extensive clinical and image interpretation experience, automatic airway segmentation can reduce the manual efforts and accelerate the reconstruction of pulmonary structures. In recent years, deep learning methods have been widely used in medical image analysis. As for automatic airway segmentation, many existing efforts \cite{media2016,meng2017,qin2019,wang2019,media2019,zhao2019} adopt convolutional neural networks (CNNs) to learn robust and discriminative features. The major challenge in this task is to accurately reconstruct complete airway tree branches, which needs a high sensitivity of distal small airways. Due to the small size and scattered spatial distribution of peripheral bronchi, the segmentation accuracy is degraded by severe class imbalance. In this paper, we demonstrate the intrinsic impact of class imbalance on the training of CNNs and propose a method to deal with this problem. 
\setcounter{figure}{0}
\begin{figure}[!t]
	\centering
	\includegraphics[scale=0.52]{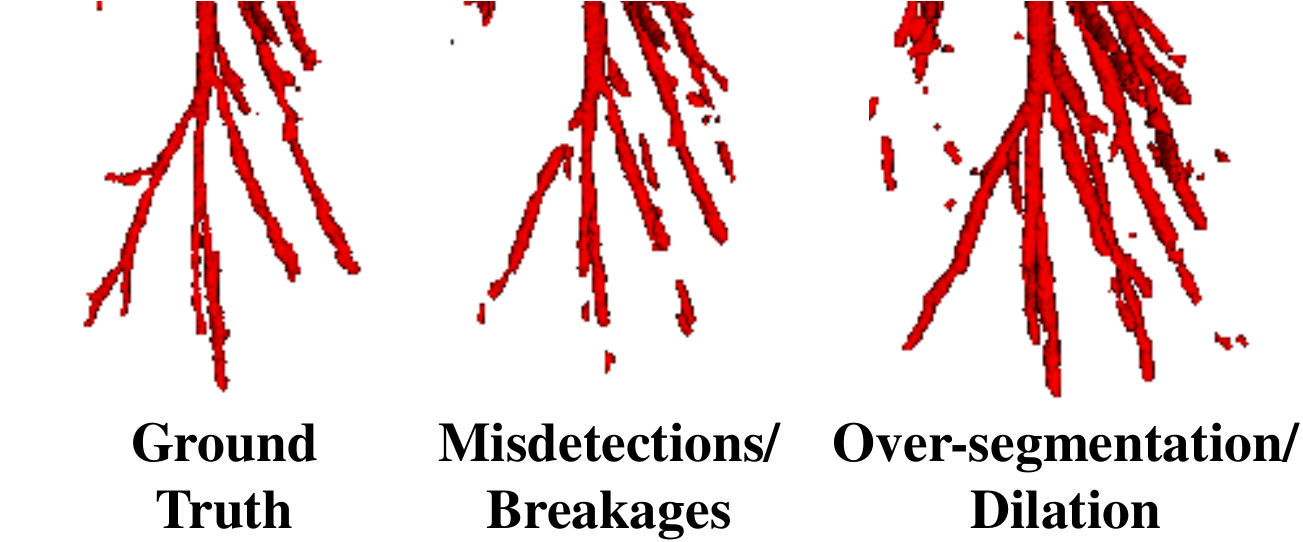}
	\caption{Illustration of the impacts of inter-class imbalance. Without larger weights for airway voxels in the loss function, the trained network may fail to segment a complete branch. In contrast, a large weight helps to detect more complete bronchi while the predicted branches are thicker than the ground truth. A part of the airway wall is misclassified as lumen.}
	\label{overseg}
\end{figure}

\begin{figure}[!t]
	\centering
	\includegraphics[scale=0.37]{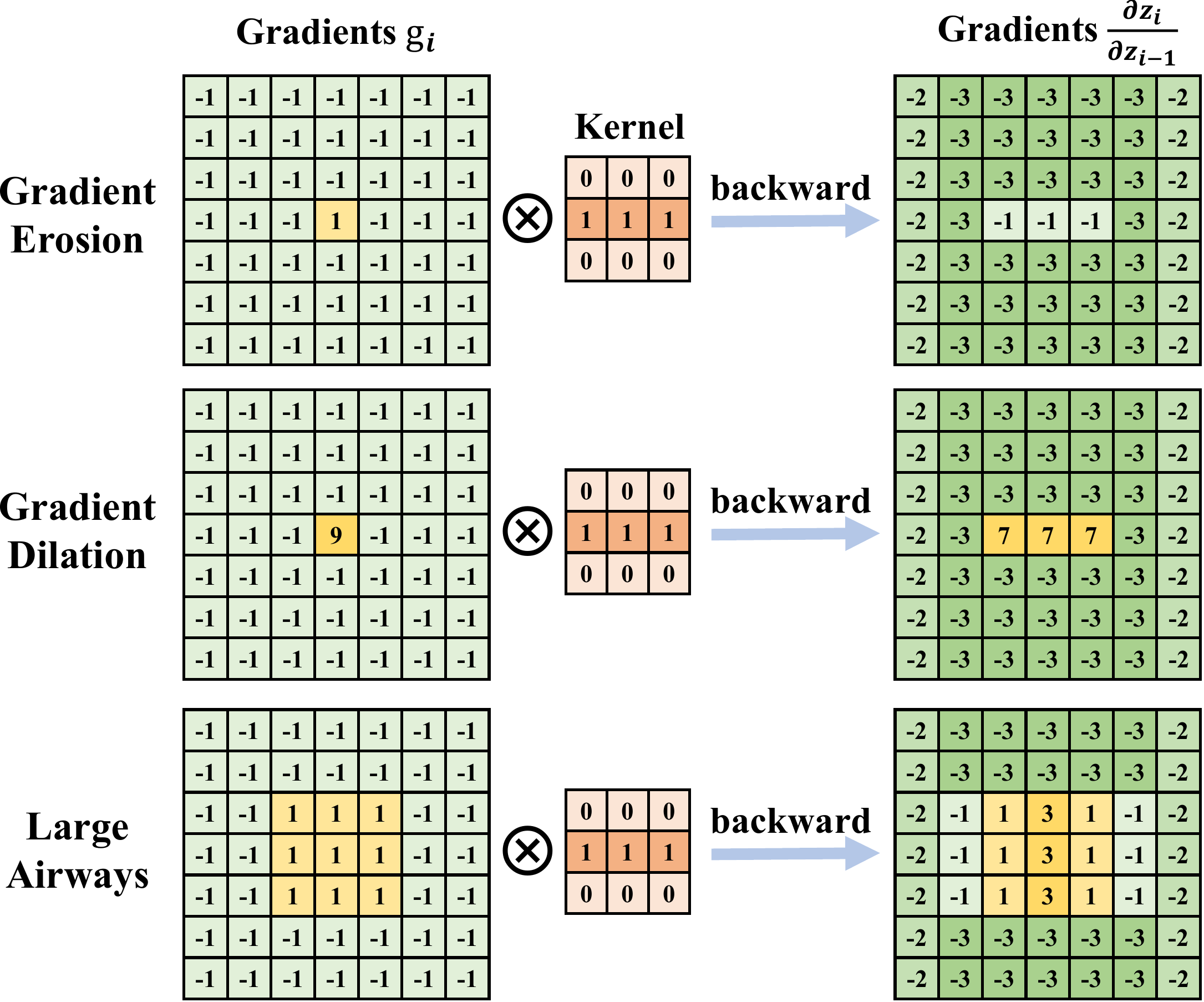}
	\caption{Illustration of gradient erosion and dilation, where $g_{i-1} = g_{i}\frac{\partial z_{i}}{\partial z_{i-1}}$ according to the chain rule. For distal airways, the small foreground gradients may be eroded by the surrounding background gradients during backpropagation, while the large foreground gradients can inversely affect their background neighborhoods. For large airways, this problem is not serious.}
	\label{gradient_ED}
\end{figure}

For airway segmentation, the first kind of class imbalance is that the number of airway voxels is far fewer than that of background, which is denoted by \textbf{inter-class imbalance} in this paper. As shown in Fig.~\ref{overseg}, if we ignore the influence of inter-class imbalance in a segmentation pipeline, the performance will be limited by misdetections and breakages of small airway branches. By assigning larger weights to the airway voxels in the loss function, more complete branches can be detected. However, their diameters can be larger than the ground truth since a part of the airway wall is misclassified as the lumen. We demonstrate that this phenomenon in airway segmentation can be interpreted as the \textbf{gradient erosion and dilation} of the neighborhood, which is closely related to the gradient ratio between foreground and background. A simple illustration is shown in Fig.~\ref{gradient_ED}. If the magnitude of the gradients to small airway points is not much larger than those of the background points, their gradients may be eroded by the surrounding background gradients via convolutional kernels. Moreover, this process is cumulative and the gradients for these branches may be thoroughly diluted when arriving at the bottom layers during backpropagation. As a result, the ``shallow layers'', which also refers to the first few layers in CNNs, do not learn to recognize these peripheral bronchi. After assigning larger weights to these points, the foreground gradients will inversely affect their background neighborhoods, leading to the dilation issue. In this case, the network tends to classify the surrounding background voxels as foreground. Such dilation may cause leakages especially at bifurcations. Although the gradient erosion and dilation also affect the large airways, the consequence is more serious for the peripheral bronchi.
Actually, Fig.~\ref{gradient_ED} is only a simple assumption while more complicated analysis can be found in the supplemental material. To improve the segmentation performance of the airway points under severe local inter-class imbalance, we can change the supervision manner to enhance the training of shallow layers.     

Another kind of class imbalance is \textbf{intra-class imbalance}, which refers to the relative total volume difference of trachea, main bronchi, lobar bronchi and distal segmental bronchi. Large airways account for the majority of the total volume, and for data-driven algorithms, such imbalanced distribution affects the segmentation performance of peripheral bronchi. In this paper, we use Dice Loss \cite{VNet} as an example to analyze the intra-class imbalance in segmentation. The Dice loss is defined as follows:
\setcounter{equation}{0}
\begin{equation}
D = 1 - \frac{2\times\sum_{i=1}^{N}p_{i}g_{i}}{\sum_{i=1}^{N}p_{i}+\sum_{i=1}^{N}g_{i}},
\end{equation} 
where $N$ is the total number of voxels, $p_{i}$ is the prediction of each voxel and $g_{i}$ is the corresponding ground truth. In this loss, the same gradients are given to all foreground $p_{f}$ or background $p_{b}$ locations ($\frac{\partial D}{\partial p_{f}} = \frac{-2\left[ \left( \sum_{i=1}^{N}p_{i}+\sum_{i=1}^{N}g_{i}\right) - \sum_{i=1}^{N}p_{i}g_{i} \right] }{\left( \sum_{i=1}^{N}p_{i}+\sum_{i=1}^{N}g_{i}\right)^{2} }$ and $\frac{\partial D}{\partial p_{b}} = \frac{2\sum_{i=1}^{N}p_{i}g_{i}}{\left( \sum_{i=1}^{N}p_{i}+\sum_{i=1}^{N}g_{i}\right)^{2} }$)\footnote {Detailed analysis can be found in the supplementary material.}. For further analysis with regard to the gradient erosion and dilation, we calculate the ratio of the foreground gradient to background gradient,
\begin{equation}
r_{Dice} = \left|\frac{\partial D}{\partial p_{f}} \left/ \frac{\partial D}{\partial p_{b}} \right. \right| = \frac{2}{1-D} - 1.
\label{dice}
\end{equation} 
This proportion is in inverse ratio to the Dice similarity coefficient (DSC). During training, if an input patch includes both small and large airways, the total DSC is high as the large airways are not seriously influenced by the inter-class imbalance. This leads to small $r_{Dice}$ which aggravates the gradient erosion, thus affecting the learning of small airways. For this problem, the design of loss function plays a vital role. 

In this paper, we address the inter-class and intra-class imbalance from the aspects of supervision manner and loss function respectively. As a major reason for gradient erosion and dilation is the successive stack of convolutional layers, in previous works, deep supervision \cite{DeepSup} adds more decoding paths after several middle layers. These supplementary supervisions provide complemental gradients for the bottom part of the network. However, within each encode-decode path, the defective information is still cumulated. To reduce this negative impact, we develop a new supervision flow named group supervision which directly transforms complemental gradients to each convolutional block (ConvBlock), which consists of a $3\times3\times3$ convolution with its corresponding normalization and activation layers. In group supervision, the ConvBlocks are divided into different groups with respective loss functions. Feature pyramid is built within each group and delivers nearly original gradients to all the members. WingsNet is built by integrating the group supervision with an UNet \cite{UNet} structure to learn the detailed representations of small targets under severe inter-class imbalance. For intra-class imbalance, it is necessary to modify the gradients of different airway voxels. On the one hand, the segmentation of thick branches does not need a large gradient ratio, which is the key to learn the representations of peripheral bronchi. On the other hand, some branches are close to the vessels or lesions, making them harder to be detected. To this end, we propose a General Union loss that assigns distance-based weights to different airway voxels and further increases the gradient ratios of the hard-to-segment regions. 

The main contributions of this work are as follows:
\begin{enumerate}
	\item Gradient erosion and dilation are first introduced to elucidate the influence of class imbalance when adopting CNNs in segmentation. Extensive experiments on two public datasets demonstrate the efficacy of our method.
	\item A novel supervision flow is developed to enhance the training of shallow layers under severe inter-class imbalance. WingsNet is built by combining the proposed group supervision with an encode-decode structure.
	\item General Union loss is proposed to deal with the intra-class imbalance via distance-based weights and element-wise focus on the hard-to-segment regions. 
\end{enumerate} 
\begin{figure*}[ht]	
	\centering
	\includegraphics[scale=0.9]{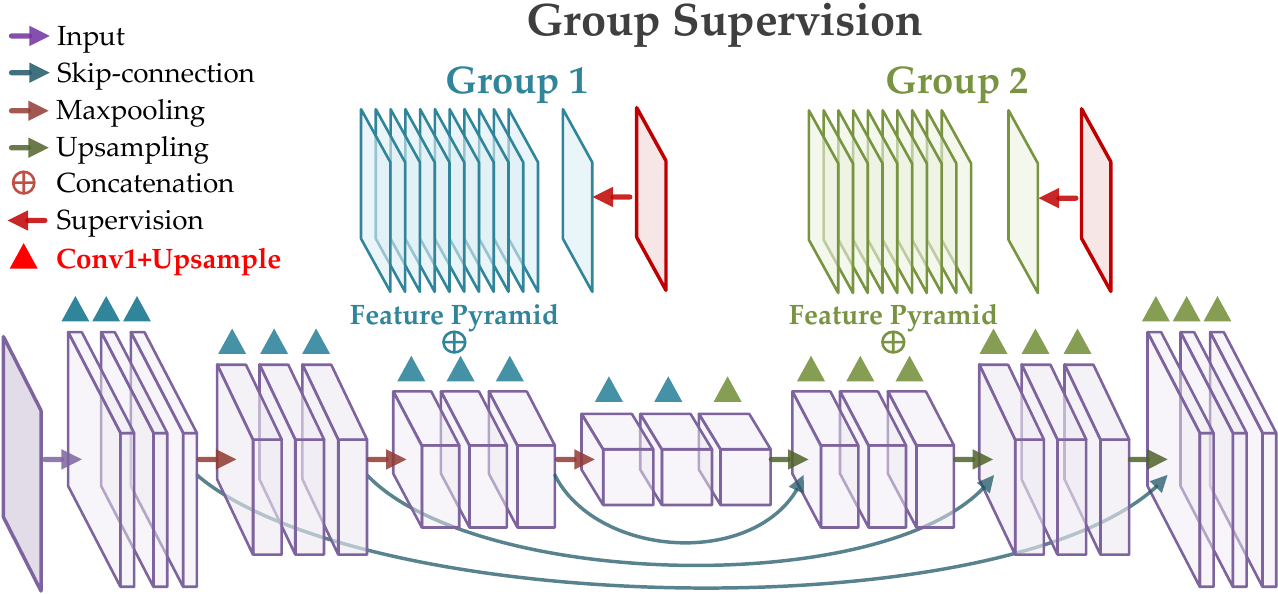}
	\caption{This figure illustrates an example of group supervision, in which the ConvBlocks (purple boxes) are divided into two groups according to the color of triangles. In each ConvBlock, there are two output paths, one to the next block and another (triangle) to the group feature pyramid. The second path consists of a $1\times1\times1$ ConvLayer (`Conv1') and an upsampling layer. Supervisions are imposed on each group.}
	\label{network}
\end{figure*}
\section{Related Work}
\subsection{Airway Segmentation}
Thus far, many methods have been proposed for airway segmentation. In 2009, fifteen algorithms (ten fully automatic and five semi-automatic) were compared in a segmentation challenge (EXACT'09 \cite{exact09}). Conventional methods such as region growing \cite{Pinho}, template matching \cite{Born} and gradient vector flow \cite{Bauer} are adopted and thick tubular structures (trachea, principle bronchi, etc.) can be well detected in these works. However, due to the lack of semantic features, the blurred or broken airway walls limit their performance in terms of peripheral bronchi.

In recent years, deep learning methods are widely used in this task \cite{media2016,meng2017,qin2019,wang2019,media2019,zhao2019,Zhang2020,mali}. Some papers directly use 3D segmentation neural networks to get the final segmentation results. Juarez \emph{et al.}\cite{Juarez2018} adopt 3D UNet as their segmentation network while Qin \emph{et al.}\cite{qin2019} propose AirwayNet which emphasizes the connectivity of voxels. Deep neural networks are also combined with conventional methods. Jin \emph{et al.}\cite{Jin2017} first train a 3D fully convolutional network (FCN) followed by a graph-based refinement. Zhao \emph{et al.}\cite{zhao2019} separately train their 3D and 2D networks, then perform linear programming to combine these results. Besides, CNNs are integrated into a tracking framework. Yun \emph{et al.}\cite{media2019} use 2.5D CNN to classify the candidates generated based on the segmentation masks in the previous iteration. Meng \emph{et al.}\cite{meng2017} track the airways along the centerlines and adopt 3D UNet to extract the airways in the volume of interest. However, the class imbalance which plays a vital role in accurate distal bronchi segmentation has not been thoroughly discussed. In this paper, we especially focus on this problem and propose the corresponding solutions.
\subsection{Class Imbalance in Image Segmentation}
The approaches to mitigate the class imbalance in segmentation can be divided into two categories, the sampling strategies and the cost-sensitive learning techniques. In 3D image segmentation, the widely used two-stage framework consisting of a localization phase and a refinement stage can be seen as an over-sampling method for the minority class. Wang \emph{et al.}\cite{wang_sample} propose a sampling policy, named Relaxed Upper Confident Bound to deal with the trade-off between exploitation and exploration in multi-organ segmentation. In the cost-sensitive learning techniques, Tversky loss \cite{tversky} and Generalized Dice loss \cite{generalized_dice} can be used to boost the sensitivity of under-represented class. Besides, focal Dice loss \cite{focalDice}, focal Tversky loss \cite{focalTversky} and exponential logarithmic loss \cite{ELloss} achieve a patch-wise focus function by changing the root of the union-based losses (Dice loss, Tversky loss, etc.). Moreover, prior knowledge in a certain task is also integrated into the loss function design \cite{wang2019}. To deal with the intra-class imbalance between large and small airways, in this paper, we incorporate the airway-tree prior knowledge as well as the element-wise focal function which assigns larger gradients to the hard-to-segment voxels.
\section{Methods}
\subsection{Overview}
Our framework is effective and robust for airway segmentation. During training, the input patches are sampled from the CT scans via a hard skeleton sampling strategy. The proposed WingsNet is built with group supervision and trained by Root Tversky loss and General Union loss. During testing, the trained WingsNet takes small patches as input by a sliding window manner. The following sections provide a detailed explanation of the group supervision, WingsNet, and General Union loss. The corresponding code is publicly available at \underline{https://github.com/haozheng-sjtu/3d-airway-segmentation}.
\subsection{Group Supervision}
\begin{figure*}[ht]	
	\centering
	\includegraphics[scale=0.75]{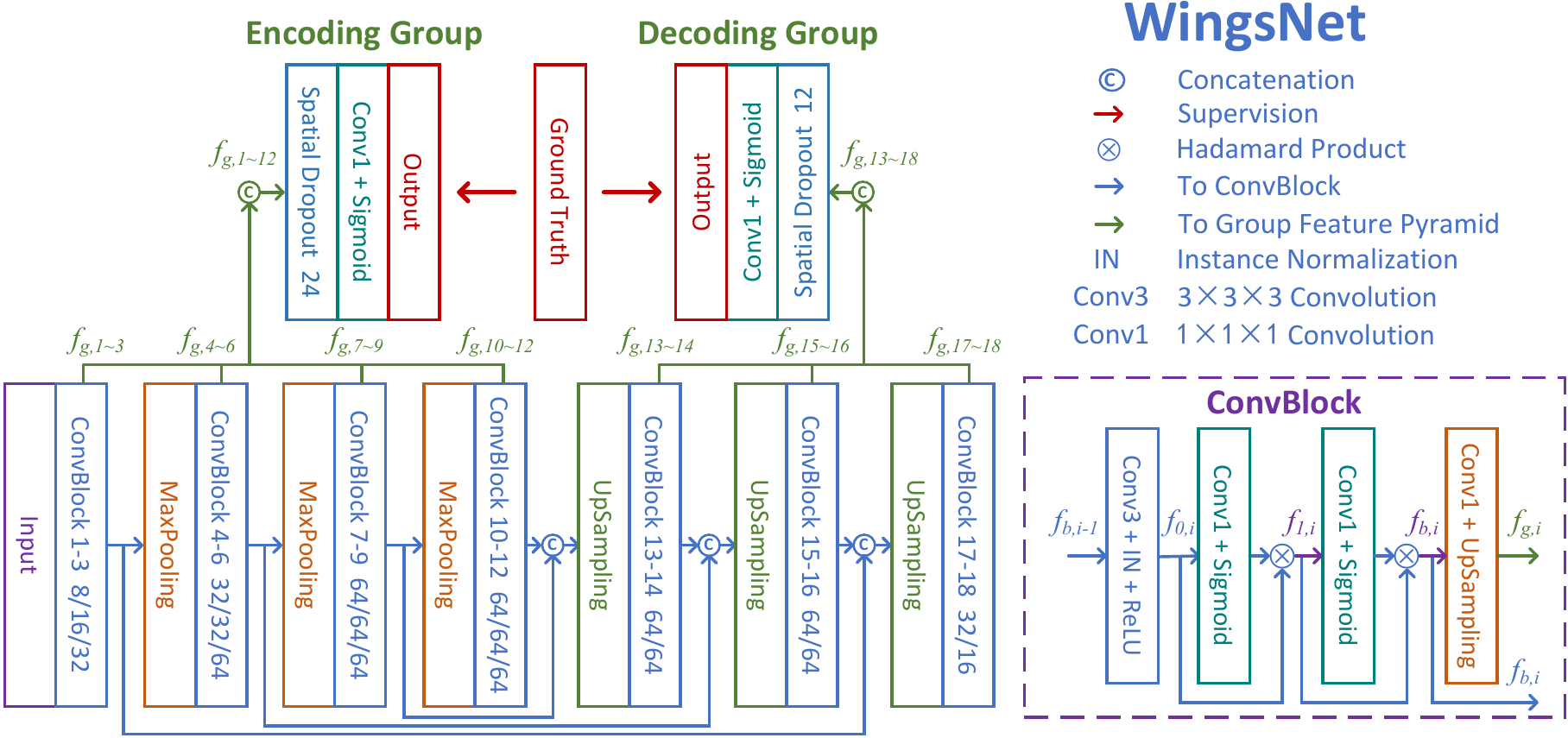}
	\caption{This figure illustrates the detailed architecture of the proposed WingNet. Within the network backbone, `Convblock 1-3 8/16/32' means the first three consecutive Convblocks with output channels of 8, 16 and 32 to their next blocks respectively. The detailed structure of each ConvBlock is also shown in the purple frame. For the $i^{th}$ ConvBlock, $f_{b,i}$ means the output to the next block and $f_{g,i}$ denotes the output to the group feature pyramid. The output channel number of each $f_{g,i}$ is two.}
	\label{wingsnet}
\end{figure*}
Under severe class imbalance, the stack of convolutional layers (ConvLayers) is a major reason for the gradient erosion and dilation problem. However, it is ineffective to deal with this by reducing the depth of the network, since this also weakens the representative ability of the model. Instead, as this problem mainly affects the training of shallow layers, we can take measures to reduce the noisy information included in the gradients when arriving at these parts. In this paper, we propose a novel supervision manner named group supervision with its corresponding network architecture called WingsNet to achieve this purpose.

In previous works, Deep supervision \cite{DeepSup} generates extra encode-decode paths while within each path, the stack of layers still leads to the gradient erosion and dilation. Deeply supervised nets \cite{DeeplySN} impose loss functions on each ConvBlock, but such dense supervision may have a side effect on the learning of hierarchical representations if the shallow and middle layers are also forced to generate a segmentation only based on their own features. To provide complemental gradients to each block while preserving the hierarchical representations, we propose to separate the ConvBlocks into different groups while supervising each of them. 
Fig.~\ref{network} demonstrates an example of group supervision. The twenty-one ConvBlocks of a UNet structure are divided into two groups according to the color of their triangles. Each ConvBlock is composed of a convolutional layer with its corresponding normalization layer and activation layer. Within each block, apart from the output path to the next block, there is an extra path to the group feature pyramid which is the concatenation of the extra outputs of its member blocks. More specifically, the output of one ConvBlock further passes through a $1\times1\times1$ convolutional layer and the size is recovered to the original size by an up-sampling layer. It should be noted that the $1\times1\times1$ ConvLayer does not contribute to gradient erosion and dilation, thus the gradients from these additional paths can help the middle layers to learn the representation of small structures. In the original UNet, the supervision is imposed on the output of the final layer. In contrast, when adopting group supervision, each group makes its own prediction based on the feature pyramid. For the $i^{th}$ group, the binary segmentation can be obtained by 
\begin{equation}
P_{i} = Sigmoid(Conv1(Cat(f_{i,1},f_{i,2},...,f_{i,n}))),
\end{equation}
where `Conv1' denotes $1\times1\times1$ convolution, `Cat' means concatenation, $f_{i,j}$ denotes the features from the $j^{th}$ block and n is the total number of blocks in this group. During inference, only the output of the last group is considered as the model prediction while the supervision on other groups is used to enhance the training. In Fig.~\ref{network}, there are two groups (blue and green). But more generally, the number of groups can be variable according to the network property. And the partition of groups is also not fixed while it is important to keep a hierarchical representation within each group. Different kinds of group supervision manners are compared in our experiment. 

To adopt the proposed group supervision for airway segmentation, we build the WingsNet with similar architecture and the details are shown in Fig.~\ref{wingsnet}. The WingsNet totally includes 18 ConvBlocks, which are divided into encoding group (ConvBlock 1-12) and decoding group (ConvBlock 13-18). For the $i^{th}$ ConvBlock, its outputs to the next block $f_{b,i}$ and the group feature pyramid $f_{g,i}$ are obtained by 
\begin{equation}
f_{0,i} = ReLU(IN(Conv3(f_{b,i-1}))),
\end{equation} 
\begin{equation}
f_{1,i} = Sigmoid(Conv1(f_{0,i}))\odot f_{0,i},
\label{SA1}
\end{equation} 
\begin{equation}
f_{b,i} = Sigmoid(Conv1(f_{1,i}))\odot f_{1,i},
\label{SA2}
\end{equation} 
\begin{equation}
f_{g,i} = Upsample(Conv1(f_{b,i})),
\end{equation}
where $f_{b,i-1}$ is the output of the previous block, $f_{0,i}$ and $f_{1,i}$ are intermediate results, `IN' is instance normalization, $\odot$ is Hadamard product, and `Conv3' means $3\times3\times3$ convolution. The illustration is in the purple frame where the four layers correspond to the four equations. As the inter-class imbalance problem stems from the large number of background points, it can be mitigated via adaptively suppressing the background locations during training. We achieved this by a spatial attention module \cite{SA}, which generates a soft mask applied on $f_{0,i}$. Moreover, we cascade two modules (Eq.~(\ref{SA1}) and Eq.~(\ref{SA2})) to generate the attention map in a coarse-to-fine manner. But limited by the capacity of GPU memory, the first three and last two ConvBlocks only include the one-stage original spatial attention module. In addition, we adopt SpatialDropout\cite{SDropout} before the last $1\times1\times1$ convolution within each group for regularization, which is first used in pose estimation and performs well in small training set\cite{SDropout}. During training, we randomly drop a certain proportion of the feature maps within each pyramid. Compared with the standard element-wise dropout, such a channel-wise approach is more suitable for the proposed group supervision. 
\subsection{General Union Loss}
In medical imaging, union-based losses such as Dice loss and Tversky loss \cite{tversky} are widely used to address the class imbalance problem. As shown in Eq.~(\ref{dice}), Dice loss can adaptively change the gradient ratio according to the segmentation performance, resulting in increased sensitivity for the minority class. However, due to the intra-class imbalance, the network trained by Dice loss may not learn to detect the distal bronchi. To improve the airway tree length detected, we can impose a lower bound to the gradient ratio, preventing the gradient erosion problem. This can be achieved by weighted Dice loss and Tversky loss.

For weighted Dice loss, 
\begin{equation}
D_{w} = 1 - \frac{2\times\sum_{i=1}^{N}w_{i}p_{i}g_{i}}{\sum_{i=1}^{N}w_{i}\left( p_{i}+g_{i}\right) },
\end{equation} 
similar to Dice loss, the same gradients are given to all foreground or background voxels ($\frac{\partial D_{w}}{\partial p_{i}}=\frac{\partial D_{w}}{\partial p_{j}}$ if $g_{i}=g_{j}$)\footnote {More details can be found in the supplementary material.}. Its gradient ratio is
\begin{equation}
\begin{aligned}
\left|\frac{\partial D_{w}}{\partial p_{f}} \left/ \frac{\partial D_{w}}{\partial p_{b}} \right. \right| = 
\frac{w_{f}}{w_{b}}(\frac{2}{1-D_{w}}-1) \geq \frac{w_{f}}{w_{b}},
\end{aligned}
\label{weighted_dice}
\end{equation}
where $w_{f}$ and $w_{b}$ are the weights for foreground and background respectively. From Eq.~(\ref{weighted_dice}), we can find three properties of weighted Dice loss. First, its gradient ratio can be adaptively changed due to $\frac{2}{1-D_{w}}-1$. Second, this change is amplified by the factor $\frac{w_{f}}{w_{b}}$. Third, the gradient ratio has a lower bound of $\frac{w_{f}}{w_{b}}$. The last two properties ensure the segmentation performance of under-represented class even when the item $\frac{2}{1-D_{w}}-1$ is seriously affected by the intra-class imbalance.
The same properties can be found in Tversky loss\cite{tversky},
\begin{equation}
T = 1 - \frac{\sum_{i=1}^{N}p_{i}g_{i}}{\sum_{i=1}^{N}(\alpha p_{i} + \beta g_{i})}.
\end{equation}   
where $\alpha + \beta = 1$. And its gradient ratio is
\begin{equation}
\begin{aligned}
\left|\frac{\partial T}{\partial p_{f}} \left/ \frac{\partial T}{\partial p_{b}} \right. \right| =\frac{1}{\alpha}\frac{1}{1-T}-1 \geq \frac{\beta}{\alpha}.
\end{aligned}
\label{tversky}
\end{equation}
By tuning the hyper-parameter $\alpha$, Tversky loss also can achieve a high sensitivity for peripheral small airways.

However, if we only use a constant lower bound to reduce the influence of intra-class imbalance for peripheral airway segmentation, the cost for a high tree length detected rate is the significant dilation problem among the large airways. To accurately segment both the large and small airways, the training is expected to adaptively pay more attention to the hard-to-segment regions such as the distal branches and the breakages. Furthermore, the gradient ratios of different airway points should vary with the branch sizes. To this end, we first propose a Root Tversky loss which integrates the voxel-level focal function into Tversky loss, and a General Union loss is generated by adding the distance-based weights to further resolve the intra-class imbalance.

Root Tversky loss achieves the element-wise focal function by changing the root of the predictions in the numerator of Tversky loss,
\begin{equation}
T_{r} = 1 - \frac{\sum_{i=1}^{N}p_{i}^{r}g_{i}}{\alpha \sum_{i}p_{i}+ \beta \sum_{i}g_{i}},
\label{r_tversky}
\end{equation}   
where $0 < r < 1$ and $\alpha + \beta = 1$. It is differentiated in term of $p_{j}$, and the gradient is
\begin{equation}
\frac{\partial T_{r}}{\partial p_{j}} = - \frac{rg_{j}p_{j}^{r-1}\left( \alpha \sum_{i}p_{i}+ \beta \sum_{i}g_{i}\right) -\alpha \sum_{i}p_{i}^{r}g_{i}}{\left(\alpha \sum_{i}p_{i}+ \beta \sum_{i}g_{i} \right) ^{2}}.
\end{equation}
As $r-1<0$, we replace $p_{i}^{r-1}$ by $(p_{i}+\epsilon_{i})^{r-1}$ where $\epsilon$ is a small positive number. The same gradients are given to all background voxels while $p_{j}^{r-1}$ controls the gradients given to foreground locations. Since $0<r<1$, much larger gradient is given to the foreground voxel with small $p_{j}$. Its gradient ratio is
\begin{equation}
\begin{aligned}
\left|\frac{\partial T_{r}}{\partial p_{f}} \left/ \frac{\partial T_{r}}{\partial p_{b}} \right. \right| = \frac{rp_{f,j}^{r-1}}{\alpha} \frac{1}{1-T_{r}}-1 \geq \frac{r}{\alpha}-1,
\end{aligned}
\end{equation}
where $p_{f,j}$ is the prediction of a foreground voxel j. Comparing  Eq.~(\ref{tversky}) and Eq.(\ref{r_tversky}), the main difference between Tversky loss and Root Tversky loss is that the amplification factor $\frac{1}{\alpha}$ in Tversky loss is replaced by $\frac{rp_{f,j}^{r-1}}{\alpha}$. In this new factor, the $p_{f,j}^{r-1}$ item achieves the voxel-wise focal function for the hard-to-segment points, while for the easy-to-segmentation airway voxels, this factor is near $\frac{r}{\alpha}$, resulting in smaller gradient ratio and alleviating the dilation problem.

When all foreground voxels are of the same importance in the loss function, due to the intra-class imbalance, the training is dominated by the majority category. To boost the segmentation performance of the minority class, we can modify the importance based on prior knowledge. To this end, the weights for each airway voxel are assigned according to their distance to the centerlines,
\begin{equation}
w_{i} = \begin{cases}
1-m\left(\frac{d_{i}}{d_{max}} \right)^{r_{d}}, &  g_{i} = 1 \\
1, & g_{i} = 0 
\end{cases}
\end{equation}
where $d_{i}$ is the shortest distance from the current location to the centerline, $d_{max}$ is the maximum $d_{i}$ in one case and $r_{d}$ controls the pattern of decay. By adding this weight to Eq.~(\ref{r_tversky}), the loss function is in the following form:
\begin{equation}
U = 1 - \frac{\sum_{i=1}^{N}w_{i}p_{i}^{r_{l}}g_{i}}{\sum_{i=1}^{N}w_{i} (\alpha p_{i}+ \beta  g_{i}) }.
\end{equation} 
To reduce the number of hyper-parameters, we set $m = \frac{1-2\alpha}{1-\alpha}$ in $w_{i}$. Its gradient with regard to $p_{j}$ is
\begin{equation}
\frac{\partial U}{\partial p_{j}} = -\frac{w_{j}rg_{j}p_{j}^{r-1}\sum_{i}w_{i}\left( \alpha p_{i}+ \beta g_{i}\right) -w_{j}\alpha \sum_{i}w_{i}p_{i}^{r}g_{i}}{[\sum_{i=1}w_{i}(\alpha p_{i}+ \beta g_{i})]^{2}}.
\end{equation}
And the gradient ratio of this loss is
\begin{equation}
\begin{aligned}
\left|\frac{\partial U}{\partial p_{f}} \left/ \frac{\partial U}{\partial p_{b}} \right. \right| &= w_{f,j}\left[ \frac{rp_{f,j}^{r-1}}{\alpha} \frac{1}{1-U}-1 \right] \\ &\geq (1-m)(\frac{r}{\alpha}-1).
\end{aligned}
\end{equation}
It is seen that the distance-based weight $w_{i}$ is directly applied to the gradient ratio of each point. As the weights of most large airway voxels are decreased, small airways play a more important role in the item $\frac{1}{1-U}$. In other words, the adaptive changing of gradient ratio depends more on the sensitivity of peripheral bronchi, thus mitigating the impact of intra-class imbalance. By combining this prior-knowledge-based weight as well as the element-wise focal function, the General Union loss aims to detect more peripheral bronchi while keeping a high accuracy for the large airways.
\section{Experiments and Results}
\subsection{Dataset}
We evaluated our method in two public datasets. The \textbf{EXACT'09} \cite{exact09} challenge provided a training set and a test set each with 20 CT scans, while no annotation is publicly available. The \textbf{Binary Airway Segmentation Dataset} \cite{airwaynet_se} included 90 CT scans (70 from LIDC \cite{LIDC} and 20 from the training set of the EXACT'09). The pixel spacing varied from 0.5 to 0.82 mm, and the slice thickness ranged from 0.5 to 1.0 mm. For the Binary Airway Segmentation Dataset, we randomly split the 90 scans into the training set, validation set, and test set with 50, 20, and 20 samples respectively. For the EXACT'09 dataset, we trained our model on the training set with the corresponding annotations from the Binary Airway Segmentation Dataset and submitted our results on the test set for evaluation. During preprocessing, the pixel values were clamped to $ \left[ -1000,600\right]$ HU, before rescaled to $ \left[ 0,255\right]$. We also masked the background voxels outside the chest with maximal intensity.
\subsection{Implementation Details}
\subsubsection{Hard Skeleton Sampling}
During training, we sampled 16 patches with a size of $ \left[ 128,128,128\right]$ from each image in each epoch via a hard skeleton sampling strategy. The network was first trained by Dice loss and the false negatives were selected as the hard-to-segment regions. In hard skeleton sampling, we randomly selected a skeleton point belonging to those regions and generated a patch containing it. We adopted the skeletonization method proposed by Lee \emph{et al.} \cite{skeleton} to extract the centerline of the airway tree. Finally, we set a threshold $p_{s}=0.5$ to determine online for random sampling or hard sample mining during training.
\subsubsection{Multi-stage Training}
In this paper, we adopted a multi-stage training strategy to boost the tree length detected. A high-sensitive segmentation was obtained in the first phase, then the network was fine-tuned to gradually increase the precision. SGD optimizer with a momentum of 0.9 and a weight decay of 0.0001 was chosen. In the first stage, the network was trained by 100 epochs. The initial learning rate was 0.01, and it was divided by 10 in the 60th and 90th epoch. During the second stage, the network was further trained by 30 epochs. The initial learning rate was 0.01, and it was divided by 10 in the 15th and 25th epoch. Besides, We performed random rotation between $[-15^{\circ}, 15^{\circ}]$ with trilinear interpolation for data augmentation. In trilinear interpolation, a threshold was used to convert the float image into the binary mask. Since a large threshold ($0.9$) could narrow down the annotation and provide a little leeway for the dilation, they were $0.7$ and $0.9$ in the first and second stages. The hyperparameters were tuned in the validation set of the Binary Airway Segmentation Dataset and experiments for each hyper-parameter are shown in the supplemental material. During the first stage, we chose $r_{l}=0.7$, $\alpha=0.1$, $r_{d}=0.5$ and $\epsilon=0.0001$ in general union loss, while in the second stage, we reset $\alpha=0.2$. 
\subsection{Segmentation Results}
\setcounter{table}{0}
\renewcommand{\thetable}{\Roman{table}}
\begin{table}[ht]
	\centering
	\caption{Comparison in the EXACT'09 dataset.} 
	\setlength{\tabcolsep}{2mm}{
		\begin{tabular}{cccc}
			\toprule[1pt]
			Method& Branch(\%)& Length(\%)& Precision(\%)\\
			\midrule[0.8pt]
			Irving \emph{et al.}\cite{Irving2009} & $43.5(19.1)$ & $36.4(17.1)$& $98.7(2.9)$\\
			Pinho \emph{et al.}\cite{Pinho} & $32.1(6.9)$ & $26.9(6.9)$& $96.6(4.9)$\\
			Bauer \emph{et al.}\cite{Bauer} & $63.0(10.4)$ & $58.4(13.2)$& $98.6(2.1)$\\
			Born \emph{et al.}\cite{Born} & $41.7(16.2)$ & $34.5(13.2)$& $\textBF{99.6(1.1)}$\\
			Feuerstein \emph{et al.}\cite{Mori09} & $76.5(13.3)$ & $73.3(13.4)$& $84.4(9.5)$\\
			Inoue \emph{et al.}\cite{Fuji} & $79.6(13.5)$ & $\textBF{79.9(12.1)}$& $88.1(13.2)$\\
			Xu \emph{et al.}\cite{XU20151} & $51.7(10.8)$& $44.5(9.4)$& $99.2(1.6)$\\
			Yun \emph{et al.}\cite{media2019} & $65.7(13.1)$ & $60.1(11.9)$& $95.4(3.7)$\\
			\midrule[0.8pt]
			Proposed & $\textBF{80.5(12.5)}$ & $79.0(11.1)$& $94.2(4.3)$\\
			\bottomrule[1pt]
	\end{tabular}}	
	\label{result_exact}
\end{table}
\begin{figure}[!t]	
	\centering
	\includegraphics[scale=0.45]{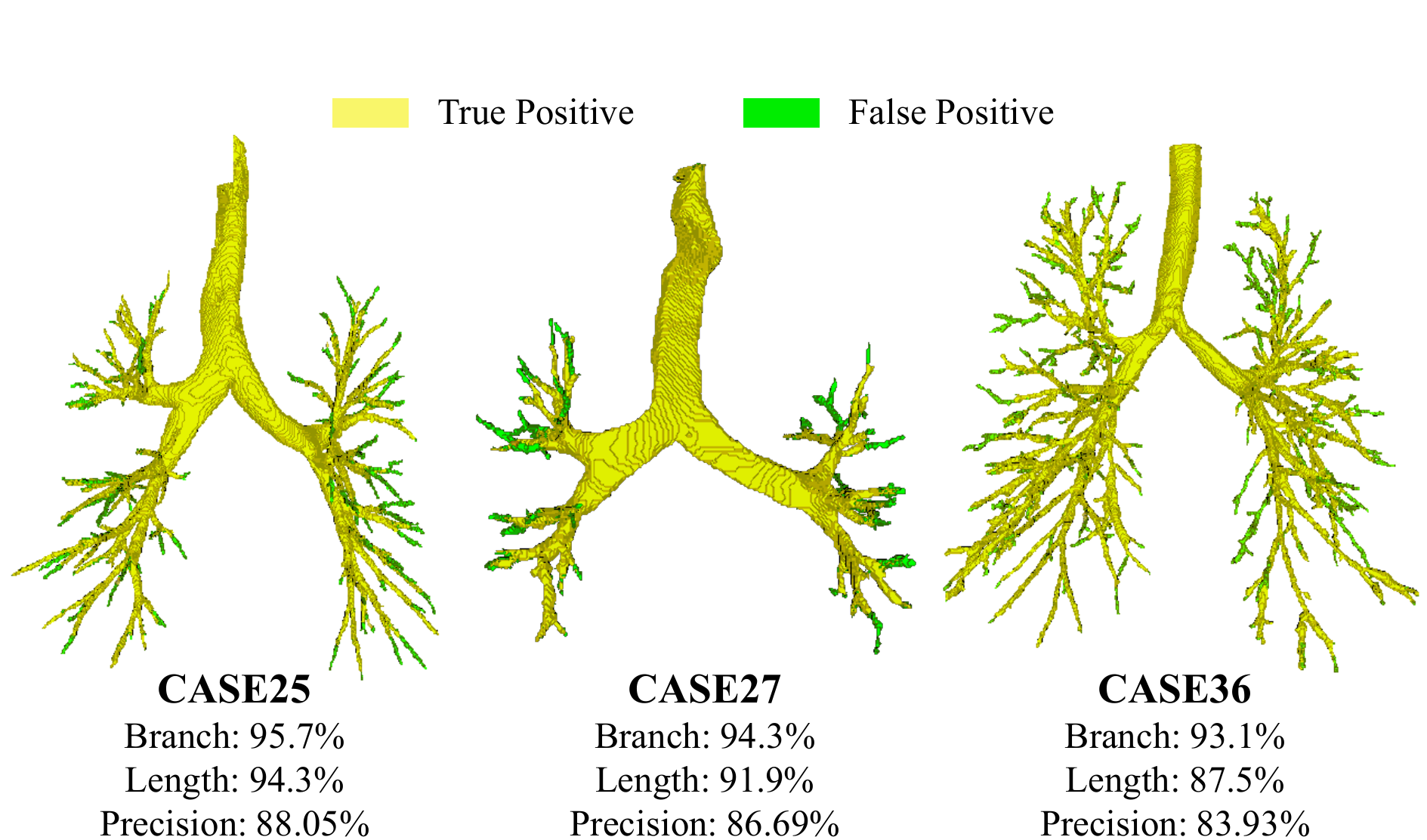}
	\caption{Three cases with the lowest precision on the EXACT'09 dataset. Most false positives belong to unannotated distal bronchi while no significant clumping leakage is observed.}
	\label{exact_results}
\end{figure}
\begin{figure*}[!t]	
	\centering
	\includegraphics[scale=0.64]{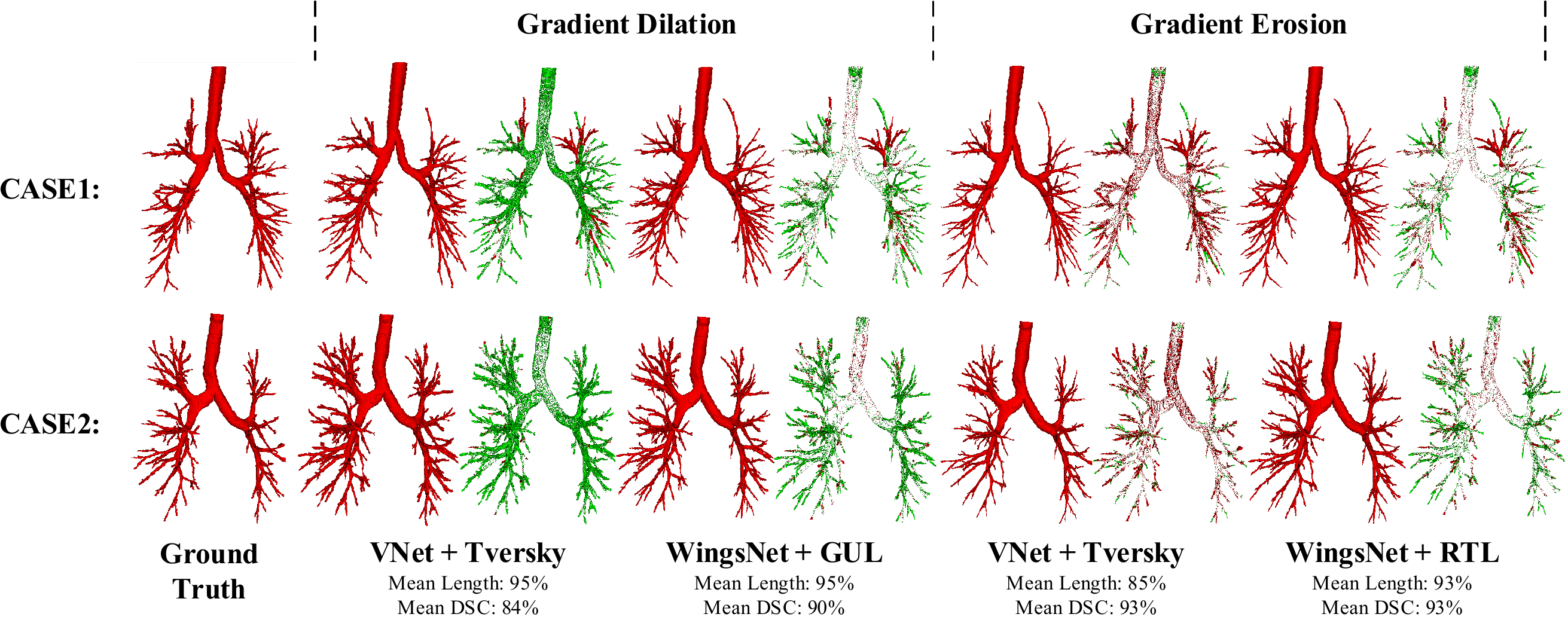}
	\caption{The segmentation results of the proposed methods and the baseline. `GUL' denotes General Union loss and `RTL' denotes Root Tversky loss. For each combination, the left figure is the segmentation result, while the right figure illustrates the false positives in green and the false negatives in red. When we used a hyper-parameter $\alpha=0.1$, with a same length detected of $95\%$, the dilation problem in the baseline is more serious. When we chose $\alpha=0.2$ and fixed the DSC to $93\%$, in the results of VNet, false negatives appear in the surface of airways as well as many distal bronchi. In contrast, the number of false negatives is much smaller in the results of WingsNet.}
	\label{results}
\end{figure*}
We adopt four evaluation metrics including tree length detected rate~\cite{Mori09}, branch detected rate~\cite{Mori09}, DSC and precision. Besides, we performed t-test for statistical analysis. Table~\ref{result_exact} illustrates the quantitative results in the EXACT'09 dataset. Among the participants of this challenge, the best branch detected ($76.5\%$) as well as length detected ($73.3\%$) rates were achieved by Feuerstein \emph{et al.}\cite{Mori09}. They adopt a tracking framework to generate the volume of interest (VOI), and a sharpening filter is used to enhance the edges of airways, followed by a region growing method to extract the branches. A major challenge in such methods is the blurry airways walls, which cause the leakages and result in a relatively low precision of $84.4\%$. Compared with their result, we achieved superior performance in all the tree metrics ($P<0.001$). In the following works on this dataset, Inoue \emph{et al.}\cite{Fuji} utilize Hessian analysis \cite{Hessian} to generate the candidates, and train an AdaBoost \cite{adaboost} classifier to reduce the false positives before obtaining the airway tree by minimum spanning tree \cite{spanning} and 3D Graph Cuts\cite {3dGraphCuts}. Compared with this method, we achieve the comparable branch detected ($80.5\% vs. 79.6\%, P>0.05$) and tree length detected ($79.0\% vs. 79.9\%, P>0.01$) with a higher precision ($94.2\% vs. 88.1\%, P<0.001$). Xu \emph{et al.}\cite{XU20151} combine a hybrid multi-scale fuzzy connectedness segmentation algorithm with two tubular structure enhancement techniques to extract the airway tree while Yun \emph{et al.}\cite{media2019} train a 2.5D CNN for bronchi segmentation. In comparison, we achieve higher sensitivity. Fig. \ref{exact_results} demonstrates three cases with the lowest precision in our segmentation results. It is seen that most false positives belong to unannotated distal bronchi while no significant clumping leakage is observed.
\begin{table}[!t]
	\centering
	\caption{Comparison in the Binary Airway Segmentation Dataset.} 
	\setlength{\tabcolsep}{1.5mm}{
		\begin{tabular}{cccc}
			\toprule[1pt]
			Method& Length(\%)& Branch(\%)& Precision(\%)\\
			\midrule[0.8pt]
			Qin \emph{et al.}\cite{qin2019} & $83.6(10.4)$& $81.4(13.8)$& $95.8(1.8)$\\
			Wang \emph{et al.}\cite{wang2019} & $86.3(8.5)$ & $83.5(11.2)$& $93.4(2.1)$\\
			Juarez \emph{et al.}\cite{Juarez2018} & $84.1(8.6)$ & $82.1(12.4)$& $91.4(2.5)$\\
			Juarez \emph{et al.}\cite{Juarez2019} & $68.0(21.1)$ & $60.5(23.9)$& $\textBF{96.4(1.8)}$\\
			Jin \emph{et al.}\cite{Jin2017} & $85.4(10.4)$ & $83.1(11.5)$& $93.9(1.9)$\\
			\midrule[0.8pt]
			Proposed & $\textBF{92.5(4.5)}$ & $\textBF{88.7(7.9)}$& $91.4(3.3)$\\
			\bottomrule[1pt]
	\end{tabular}}	
	\label{result_t}
\end{table}

Table.~\ref{result_t} compares our results with other CNN-based methods on the Binary Airway Segmentation dataset. Qin \emph{et al.}\cite{qin2019} develop AirwayNet which predicts the connectivity of airways voxels. The prediction of connectivity improves the precision but the sensitivity is still limited since it does not explicitly mitigate the gradient erosion problem. Wang \emph{et al.}\cite{wang2019} train their spatial fully connected network with radial distance loss, which places higher weights to the centerlines while reducing the weights of airway voxels far away from the centerlines, alleviating the intra-class imbalance and resulting in higher length detected ($86.25\%$). Juarez \emph{et al.}\cite{Juarez2018} train a 3D UNet with Dice loss and weighted binary cross entropy (wBCE) loss. The wBCE loss gives higher weights to the airway voxels, increasing the sensitivity. They also design a UNet-GNN architecture \cite{Juarez2019} by replacing the two convolutional layers in the deepest level of the 3D UNet by a Graph Neural Network module. In our experiment, we found that the GNN layers failed in several cases, resulting in a much higher standard deviation ($21.1\%$) than other methods. Jin \emph{et al.}\cite{Jin2017} train a 3D UNet with wBCE loss for coarse segmentation followed by graph-based refinement. The refinement contributes to both the length detected and precision. Compared with these methods, our approach can improve the tree length detected more than $6\%$ only with a little decrease in precision. Fig.~\ref{results} visualizes the segmentation results of the proposed method as well as a baseline using VNet trained by Tversky loss. The gradient ratios were tuned by choosing different hyper-parameters ($\alpha=0.1$ or $0.2$) in the loss function. First, from the results of the baseline, it is seen that the gradient erosion and dilation can appear in both large and small airways, but the consequence is more serious for the distal branches. Besides, with a large gradient ratio, when the over-segmentation problem occurs in the entire airway tree in the baseline, the dilation around the large airways is mitigated by the proposed General Union loss. Under a reduced gradient ratio, the baseline method suffers the gradient erosion problem, leading to more undetected peripheral branches. After performing group supervision, the erosion around large airways becomes insignificant while slight dilation is observed in surface of segmental bronchi.

\begin{figure*}[ht]	
	\centering
	\includegraphics[scale=0.63]{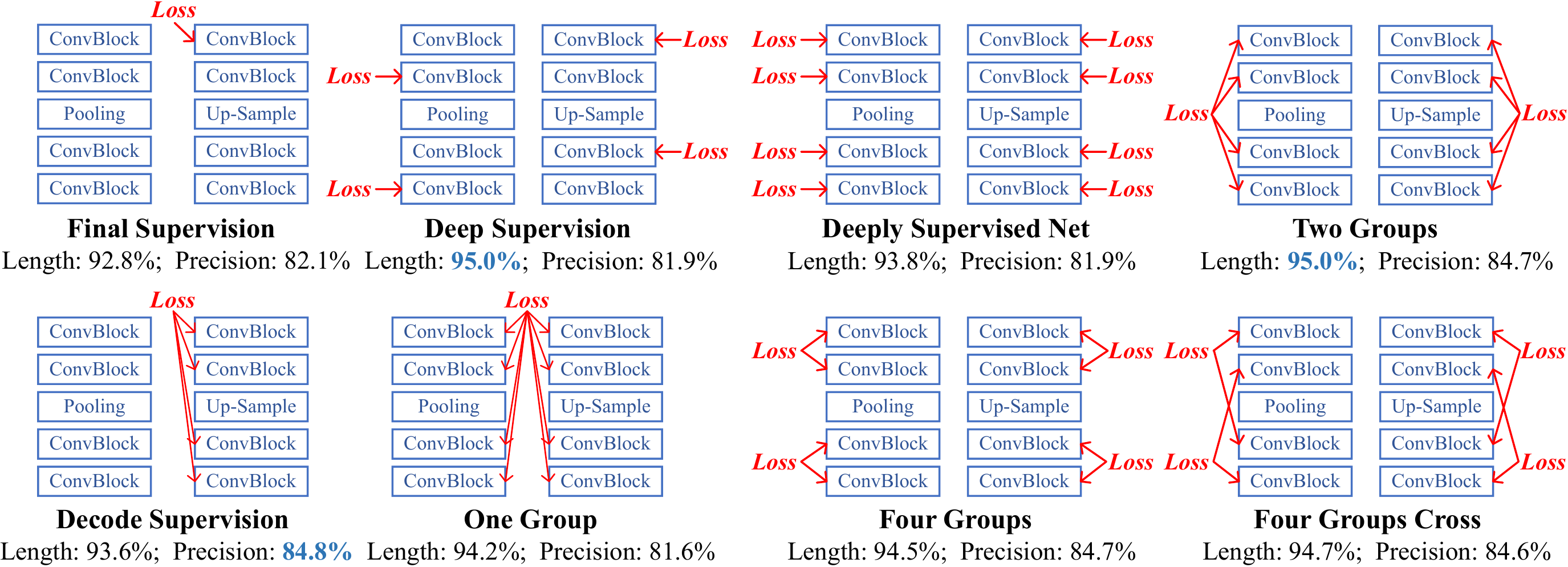}
	\caption{Comparison of different supervision approaches.}
	\label{supManner}
\end{figure*}
We further compare different supervision methods under the scenario of a high tree length detected rate ($\alpha=0.1$). As shown in Fig.~\ref{supManner}, adopting deep supervision can improve the length detected rate from $92.8\%$ to $95\%$ ($P<0.001$) while keeping a comparable precision ($P>0.05$). However, supervising each ConvBlock leads to a $1.2\%$ decrease in length detected. By dividing the blocks into two groups, with the same length detected of $95\%$, the precision is increased by $2.8\%$ ($P<0.001$). When the supervision on the encoding path is removed, the length detected drops about $1.4\%$ ($P<0.001$). Besides, if all the blocks are divided into one group, the performance is even worse than deep supervision. In contrast, adding the group number to four does not affect the segmentation results significantly. Moreover, for the four groups, we separated them via a successive manner as well as a cross manner. However, no significant difference is observed in the experiment. By these comparisons, it is shown that the segmentation of small structures benefits from the supplemental gradients while it is important to keep a hierarchical representation within each group.
\begin{figure*}[ht]	
	\centering
	\includegraphics[scale=0.53]{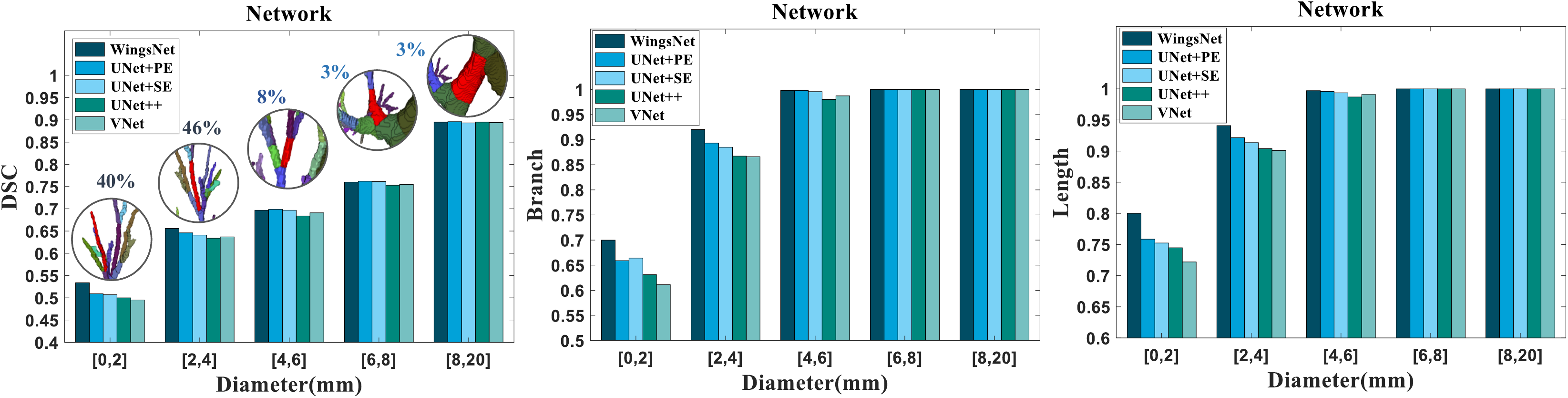}
	\caption{The segmentation results of five networks within different airway diameter intervals. The round pictures give examples for the corresponding diameter and the percentage of branches within each interval is also labeled.}
	\label{ablation_network}
\end{figure*}
\begin{figure*}[ht]	
	\centering
	\includegraphics[scale=0.53]{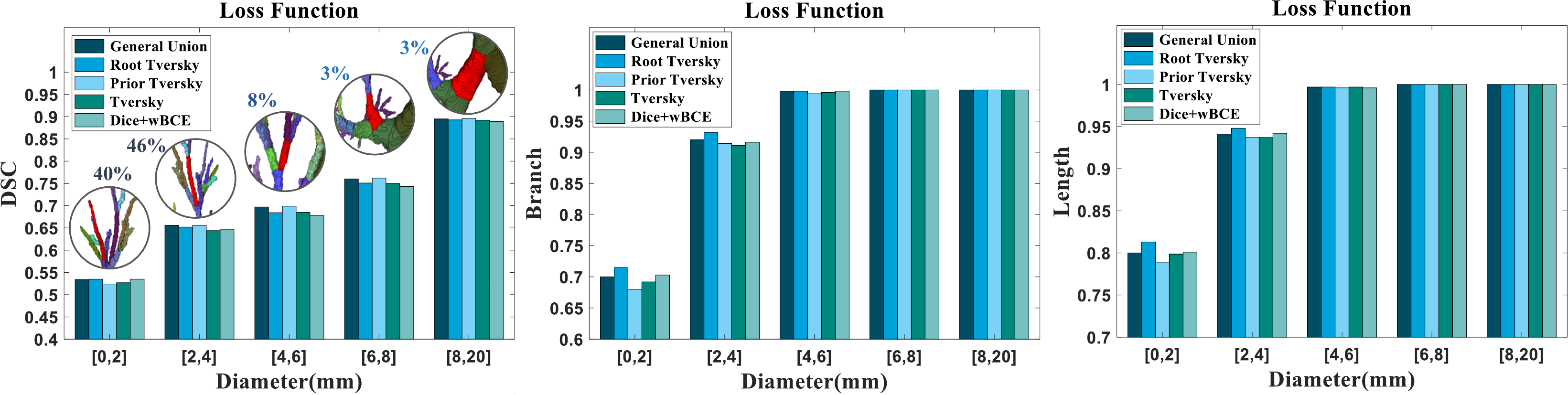}
	\caption{The segmentation results of WingsNet trained by five loss functions. The round pictures give examples for the corresponding diameter in which the selected branch is colored in red. The percentage of branches within each interval is also labeled.}
	\label{ablation_loss}
\end{figure*}
\subsection{Ablation Study}
\begin{table}[!t]
	\centering
	\caption{Ablation experiment about three key components in our segmentation framework. `GS', `GUL' and `HSS' are the abbreviations for Group Supervision, General Union Loss and Hard Skeleton Sampling.} 
	\setlength{\tabcolsep}{1.5mm}{
		\begin{tabular}{cccccc}
			\toprule[1pt]
			GS&GUL&HSS& Length(\%)& Branch(\%)& Precision(\%)\\
			\midrule[0.8pt]
			$\times$&$\times$&$\times$& $89.2(6.6)$& $84.5(9.2)$& $90.0(3.6)$\\
			$\checkmark$&$\times$&$\times$& $90.5(6.4)$& $85.9(8.7)$& $90.0(3.6)$\\
			$\times$&$\checkmark$&$\times$& $89.5(6.2)$& $84.1(9.3)$& $93.2(2.7)$\\
			$\times$&$\times$&$\checkmark$& $90.8(5.8)$& $87.1(9.0)$& $89.5(3.6)$\\
			$\checkmark$&$\checkmark$&$\times$& $90.7(6.6)$& $86.8(9.3)$& $92.5(2.7)$\\
			$\checkmark$&$\times$&$\checkmark$& $91.8(5.5)$& $87.9(8.6)$& $89.4(3.8)$\\
			$\times$&$\checkmark$&$\checkmark$& $90.2(6.0)$& $86.2(9.1)$& $92.4(2.9)$\\
			$\checkmark$&$\checkmark$&$\checkmark$& $92.5(4.5)$& $88.7(7.9)$& $91.4(3.3)$\\
			\bottomrule[1pt]
	\end{tabular}}	
	\label{ablation_study}
\end{table}
To further analyze the key components in our segmentation framework, we chose a baseline using final supervision, Tversky loss, and random sampling. The quantitative results are shown in Table.~\ref{ablation_study}. By adopting group supervision, both the length and branch detected rates are improved since the gradient erosion problem is alleviated. In contrast, the General Union loss mainly contributes to the precision. Compared with random sampling, the hard skeleton sampling helps the training focus more on the hard-to-segment peripheral regions, resulting in higher sensitivity but a slight decrease in precision. GUL and HSS can be combined with GS to further boost the precision and sensitivity respectively. But the combination of GUL and HSS leads to a trade-off between recall and precision. Comparing `GS+GUL+HSS' with `GS+HSS', it is seen that the GUL also could improve the tree length and branch detected rates by mitigating the intra-class imbalance.

To compare different network architectures, we calculated the DSC, branch detected rate, and length detected rate within five airway diameter intervals. Deep supervision is inserted into all the baselines~\cite{VNet,PE,SA,UNet++} while spatial attention is adopted by `UNet+PE'~\cite{PE} and `UNet+SE'~\cite{SA}. As shown in Fig.~\ref{ablation_network}, all the networks can achieve a high DSC for the thicks airways. However, in the test set of Binary Airway Segmentation dataset, the branches with a diameter larger than $4mm$ only account for $14\%$ of the total number. The remaining $76\%$ of branches belong to the small airways whereas the segmentation performance drops significantly for this class. The result of our WingsNet is higher than the baselines in all the three metrics within the interval between 0 and $2mm$, demonstrating the efficacy of group supervision for peripheral bronchi segmentation. We also compared five loss functions in the same manner and the results are shown in Fig.~\ref{ablation_loss}. Compared with General Union loss, Root Tversky loss does not explicitly decrease the weights of large airways, leading to a dilated segmentation and a decreased DSC for the branches with a diameter between $2mm$ and $8mm$. However, since larger weights are assigned to all voxels, the branch and length detected rates of RTL is higher than that of GUL. In contrast, without the voxel-wise focal function, Prior Tversky loss shows lower DSC and sensitivity in terms of the distal small airways. The combination of Dice loss and wBCE loss can achieve a comparable performance for the peripheral bronchi. But meanwhile, the DSC is decreased for the other branches due to the dilation problem.
\section{Discussion}
\subsection{False Positives}
\begin{figure}[ht]	
	\centering
	\includegraphics[scale=0.36]{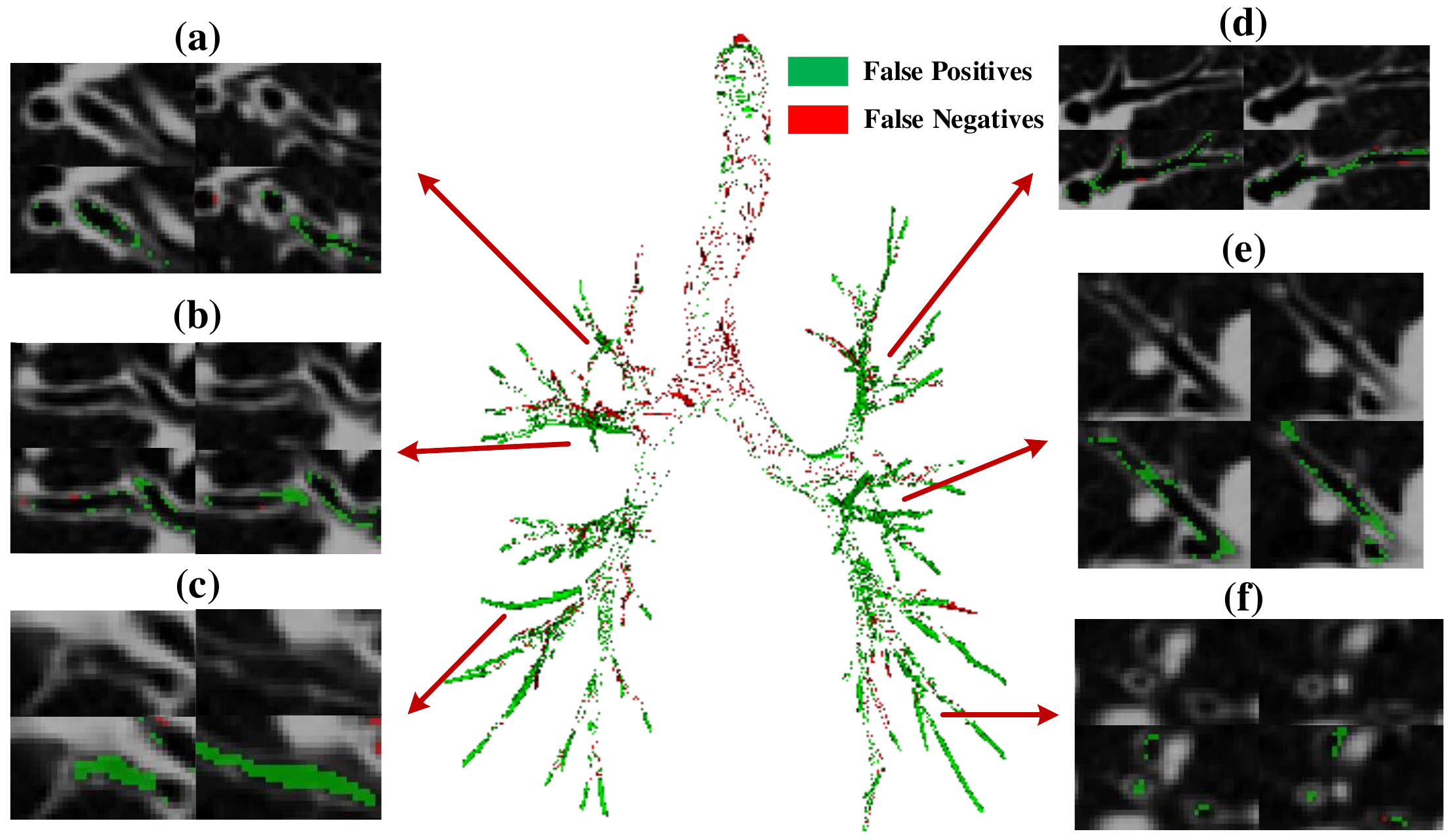}
	\caption{False positives in the segmentation results. We select six significant regions with false positives and demonstrate the original CT images in the first row with the analysis images in the second row. In the analysis images, the false positives are colored in green and the false negatives are colored in red. The same color map is used in the 3D view.}
	\label{FPs}
\end{figure}
The main drawback of our method is the decline in precision. Compared with some approaches, although the tree length detected improves about $6\%$, the decrease in precision is also more than $2\%$. In our segmentation results, the False Positives (FPs) can be divided into three categories and Fig. \ref{FPs}  illustrates six regions including FPs with their original CT images. As shown in sub-figures (a) (b) (d) (e), the largest number of FPs appear in the indistinct borders between airway lumen and airway walls. During the annotation process, observers refined the coarse segmentation obtained by the region growing method. To reduce the leakages, we set the region growing particularly sensitive to these indistinct regions. As a result, such blurry areas were not delineated in the trachea and large bronchi. Moreover, the manual annotation could not be perfect for such complicated tree structures and some indistinct regions were annotated to keep the connectivity. During training, these points were emphasized by the focal loss, and their patterns were learned by the network. Besides, the gradient dilation problem can cause the over-segmentation. As for the second class of FPs, a small number of the distal small airways with indistinct airway walls were not annotated by the observers, while some of them were detected by our method. The third class is composed of the leakages especially at the bifurcations where the intensities are between airway lumen and walls, as shown in sub-figure (e). In general, the dilation issue in the final segmentation is not as significant as that in the first stage, while the first class of FPs do not affect the clinical use. 
\subsection{False Negatives}
\begin{figure}[ht]	
	\centering
	\includegraphics[scale=0.36]{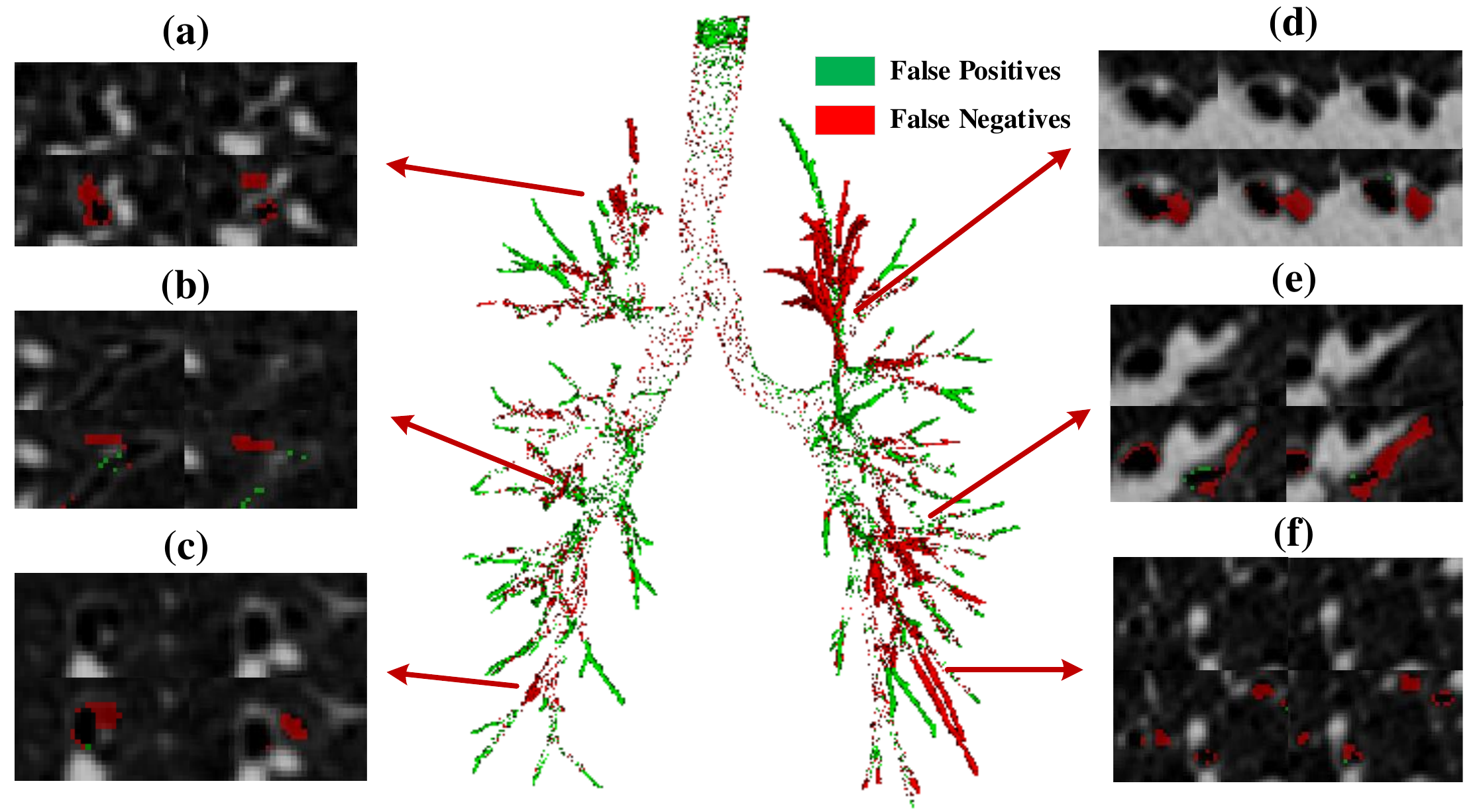}
	\caption{False negatives in the segmentation results. We select six significant regions with false negatives and demonstrate the original CT images in the first row with the analysis images in the second row. In the analysis images, the false positives are colored in green and the false negatives are colored in red. The same color map is used in the 3D view.}
	\label{FNs}
\end{figure}
We illustrate the false negatives of the failure case with the shortest length detected in Fig.~\ref{FNs}. Generally, there are two kinds of FNs. The first category can be seen in the indistinct borders between airway lumen and walls. These blurry regions are caused by the limited resolution of CT scans and the FNs in these areas do not affect the clinical use. The second class is composed of undetected branches. As shown in the sub-figures, a common feature of these undetected bronchi is that the intensity distribution of their airway walls is inhomogeneous. In other words, a fraction of the walls is of high intensity while other parts are relatively blurry. These FNs can lead to breakages, and after extracting the largest connected domain, the rest of a branch is discarded. The network failed to recognize these parts since such patterns was much common in the background. But this also reflects that the network does not learn sufficient long-range features for the connectivity of thin bronchi.
\subsection{Clinical Potentials}
The main clinical usage of airway segmentation includes disease assessment as well as bronchoscopic navigation. The measurements of bronchial morphometric parameters such as total airway count, wall thickness, and lumen diameter can be used in the diagnosis of chronic obstructive pulmonary disease (COPD) \cite{COPD}, cystic fibrosis \cite{CysticFibrosis} and asthma \cite{asthma}. Compared with previous approaches, the proposed method improves the segmentation performance of peripheral bronchi, providing a more precise count of airway branches. Although the precision drops due to the gradient dilation problem, the false positives located at the transitional areas between airway lumen and walls do not significantly affect the measurements of wall thickness as well as internal diameter. 

3D navigated bronchoscopy plays an increasingly important role in the diagnosis and treatment of lung cancer, especially for the potentially curable lesions, which are often more peripherally located \cite{Bronchoscopy}. Although thinner bronchoscopes are being developed continuously, it is still challenging to reach the peripheral targets due to complicated airway tree structures. In this situation, detailed segmentation of the bronchial tree is needed to build the virtual lung model for path planning and navigation. At present, most commercially available navigation software cannot accurately segment small distal airways, which needs to be delineated manually. However, the manual delineation of peripheral bronchi is laborious and time-consuming, an automatic segmentation for more peripheral branches is essential to accelerate the preoperative process. Moreover, such a method also makes it practical to statistically analyze the anatomy of peripheral airways with large population and serial examination data, which is helpful for the pulmonologists and can be further used to guide the segmentation.
\section{Conclusion}
In this paper, we focus on both the inter and intra-class imbalance problems which limit the efficacy of CNNs for distal airway segmentation. With increasing emphasis on early detection and endobronchial intervention, the requirement for complete airway reconstruction at the resolution limit is becoming increasingly important clinically. There is thus far no effective techniques that can address such needs. To deal with the inter-class imbalance between the foreground and background, we propose the gradient erosion and dilation problem and design a group supervision manner to enhance the training of the network. To resolve the intra-class imbalance between large and small airways, we propose a General Union loss that obviates the impact of airway size by distance-based weights and tunes the gradient ratios of each airway voxel based on the learning process. The proposed method helps to detect more peripheral bronchi, leading to a viable clinical tool for small airway reconstruction and navigation. 
\section{Supplementary Materials}
\subsection{Gradient Erosion and Dilation}
\begin{figure*}[!t]	
	\centering
	\includegraphics[scale=0.85]{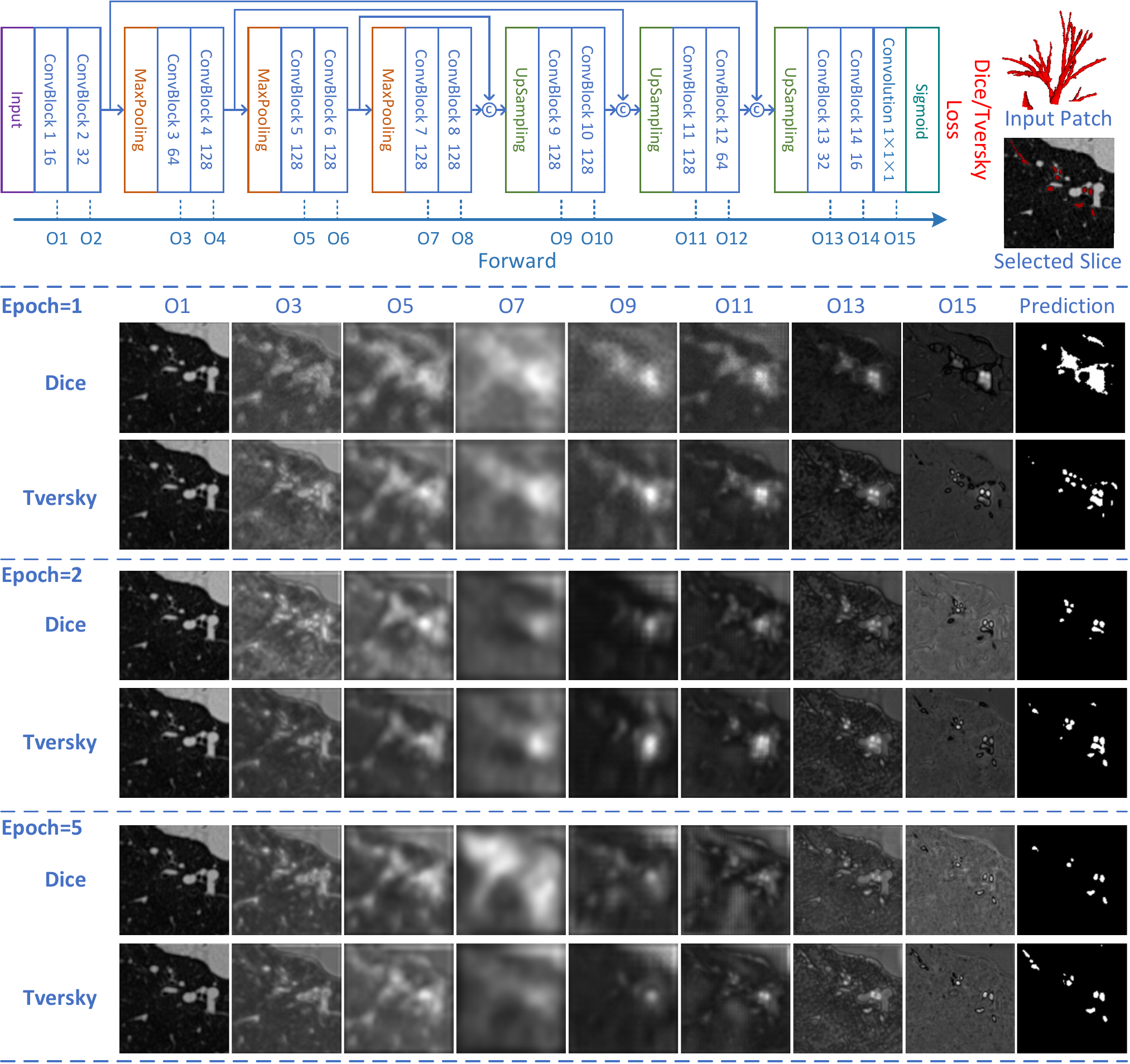}
	\caption{The intermediate outputs in a simple UNet trained by Dice loss or Tversky loss ($\alpha=0.2$). The detailed structure of the UNet is shown in the upper part. Each ConvBlock consists of a $3\times3\times3$ convolutional layer, an instance normalization layer and a ReLU layer. The attention of the outputs of eight selected layers after 3 epochs is illustrated in the lower part. }
	\label{GEAD_outputs}
\end{figure*}

This work started from the observation that the network trained by Dice loss failed to achieve a satisfactory sensitivity for small airways, whereas after assigning larger weights to the airway points, the prediction of the branches was thicker than the ground truth. To figure out the breakages occur from which layer, we checked the intermediate output of each convolutional layer in a simple UNet structure. The output attention of layer m is obtained by
\begin{equation}
O^{m} = \sum_{c=1}^{N}\left| f_{c}^{m}\right|,
\end{equation} \\
where $f_{c}^{m}$ means the $c^{th}$ channel in the features from layer m, N is the total channel number of this feature map. $O^{m}$ was further rescaled to $\left[ 0,1\right] $ by the maximum. Fig.~\ref{GEAD_outputs} illustrates the detailed structure of the used UNet as well as the output attention maps of 8 selected layers. The UNet totally included 14 ConvBlocks. Each block consisted of a $3\times3\times3$ convolutional layer, an instance normalization layer and a ReLU layer. For an input patch mainly including small airways, we compared the intermediate results of Dice loss and Tversky loss ($\alpha=0.2$) after the first, second and fifth epochs. It is seen that after the fifth epoch, a major difference is that the branch in the upper left corner can be detected in the result of Tversky loss but is missed in that of Dice loss. From the intermediate outputs of Tversky loss, it is further found that at the beginning of the decoding ($O^{9}$), this branch is not noticed. After concatenating the low-level features from skip connection, $O^{11}$ and $O^{13}$ gradually pay attention to this area. Therefore, we believe that the low-level features play an important role in the detection of peripheral bronchi. 

\begin{figure*}[!t]	
	\centering
	\includegraphics[scale=0.85]{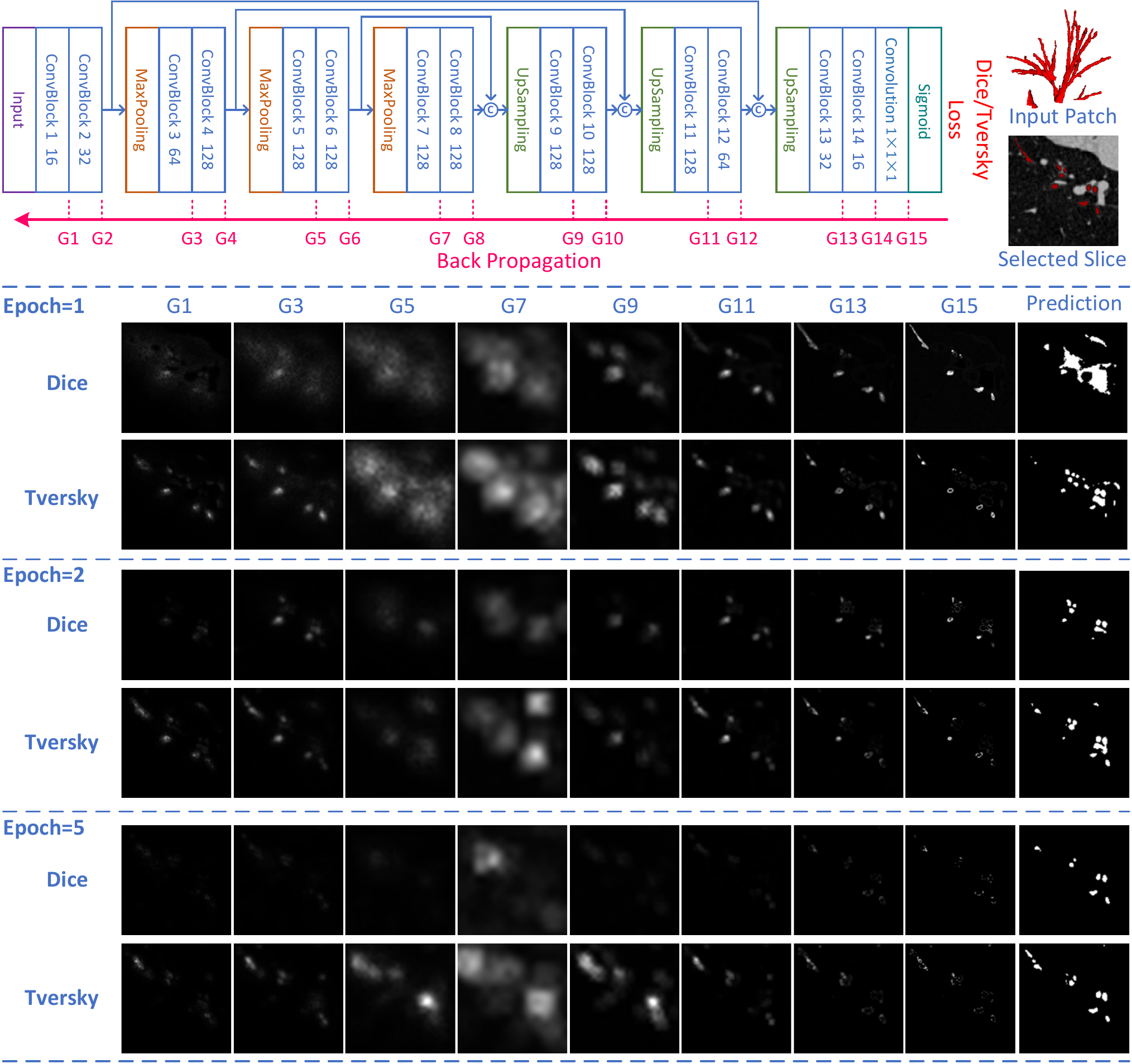}
	\caption{The intermediate gradients in a simple UNet trained by Dice loss or Tversky loss ($\alpha=0.2$). The detailed structure of the UNet is shown in the upper part. Each ConvBlock consists of a $3\times3\times3$ convolutional layer, an instance normalization layer and a ReLU layer. The attention of the gradients to eight selected layers after 3 epochs is illustrated in the lower part. }
	\label{GEAD_grads}
\end{figure*}

To further analyze the training of shallow layers, we checked the gradient flow to each convolutional layer. The gradient attention of layer m is defined as
\begin{equation}
G^{m} = \sum_{c=1}^{N}\left| g_{c}^{m}\right|,
\end{equation} \\
where $g_{c}^{m}$ denotes the $c^{th}$ channel in the gradient to layer m, N is the total channel number of this gradient map. We also demonstrates the gradient attention maps of the 8 selected layers in the same three epochs. As shown in Fig.~\ref{GEAD_grads}, in the first epoch of Dice loss, the gradient to this branch is significant in the last few layers. However, when arriving the shallow layers, the gradient to this area has been diluted. In the following epochs, due to the Sigmoid layer ($S'(x)=S(x)\times(1-S(x))$), the gradient to this branch is filtered out at the beginning of backpropagation. In contrast, during the training with Tversky loss, the gradient attention to this region can reach the bottom layers. As a result, these kernels can keep learning the representation of this branch. Moreover, the attention is also given to the surrounding areas of this branch, and as shown in Fig.~\ref{GEAD_outputs}, during the inference of the model trained by Tversky loss, the area of the output attention to each airway branch is larger than that of the Dice loss, leading to a dilated segmentation result.

Fig. 6 in the paper also demonstrates the results of gradient erosion and dilation. From the results of `VNet+Tversky', it is seen that with a small gradient ratio, the erosion appears in both large and small airways, while the consequence is more serious for the peripheral bronchi. Similarly, with a large gradient ratio, the over-segmentation problem occurs in all branches. After fine-tuning the gradient ratios of different airway voxels in General Union loss, a more accurate segmentation of both large and small airways is obtained. This demonstrates that the gradient erosion and dilation problem is highly related to the gradient ratio.

Apart from fine-tuning the gradient ratios, we also tried to resolve the gradient erosion problem by the supervision manner and finally proposed the group supervision. With the same backbone of WingsNet, if we use final supervision, to achieve a length detected of $95\%$, we need an $\alpha=0.06$ in General Union loss. In contrast, after adopting the group supervision, the same length detected rate can be achieve by $\alpha=0.1$. In other words, the suitable gradients arrive at the bottom layers under a smaller gradient ratio.\\
In summary, the gradient erosion and dilation are intrinsic properties of the successive stack of convolutional layers, which are closely related to the gradient ratio between foreground and background. However, the consequence is more serious for the voxels with severe local class imbalance. 
\subsection{Loss Functions}
In this section, we provide more detailed analysis about the loss functions used in this paper.

The Dice loss~\cite{VNet} is defined as follows:
\begin{equation}
D = 1 - \frac{2\times\sum_{i=1}^{N}p_{i}g_{i}}{\sum_{i=1}^{N}p_{i}+\sum_{i=1}^{N}g_{i}},
\end{equation} 
where $N$ is the total number of voxels, $p_{i}$ is the prediction of each voxel and $g_{i}$ is the corresponding ground truth. It is differentiated in terms of $p_{j}$, and the gradient is
\begin{equation}
\frac{\partial D}{\partial p_{j}} = - 2\left[ \frac{g_{j}\left( \sum_{i=1}^{N}p_{i}+\sum_{i=1}^{N}g_{i}\right) - \sum_{i=1}^{N}p_{i}g_{i} }{\left( \sum_{i=1}^{N}p_{i}+\sum_{i=1}^{N}g_{i}\right)^{2} }\right].
\end{equation}
It can be seen that the same gradients are given to all foreground or background locations ($\frac{\partial D}{\partial p_{i}}=\frac{\partial D}{\partial p_{j}}$ if $g_{i}=g_{j}$).For further analysis with regard to the gradient erosion and dilation, we calculate the ratio of the foreground gradient $\frac{\partial D}{\partial p_{f}}$ to background gradient $\frac{\partial D}{\partial p_{b}}$,
\begin{equation}
\begin{aligned}
r_{Dice} = \left|\frac{\partial D}{\partial p_{f}} \left/ \frac{\partial D}{\partial p_{b}} \right. \right| &= \frac{\sum_{i}\left(p_{i}+g_{i}\right)-\sum_{i}p_{i}g_{i}}{\sum_{i}p_{i}g_{i}}\\ &= \frac{2}{1-D} - 1.
\end{aligned}
\end{equation} 
This proportion is in inverse ratio to the Dice similarity coefficient (DSC). During training, if an input patch includes both small and large airways, the total DSC is high as the large airways are not seriously influenced by the inter-class imbalance. This leads to small $r_{Dice}$ which aggravates the gradient erosion, thus affecting the learning of small airways. To improve the airway tree length detected, we can impose a lower bound to the gradient ratio, preventing the gradient erosion problem. This can be achieved by weighted Dice loss and Tversky loss~\cite{tversky}.

For weighted Dice loss, 
\begin{equation}
D_{w} = 1 - \frac{2\times\sum_{i=1}^{N}w_{i}p_{i}g_{i}}{\sum_{i=1}^{N}w_{i}\left( p_{i}+g_{i}\right) },
\end{equation} 
it is differentiated in terms of $p_{j}$, and the gradient is 
\begin{equation}
\frac{\partial D_{w}}{\partial p_{j}} = -2\frac{w_{j}g_{j}[\sum_{i=1}w_{i}(p_{i}+g_{i})]-w_{j}\sum_{i}w_{i}p_{i}g_{i}}{[\sum_{i=1}w_{i}(p_{i}+g_{i})]^{2}}.
\end{equation}
Similar to Dice loss, the same gradients are given to all foreground or background voxels ($\frac{\partial D_{w}}{\partial p_{i}}=\frac{\partial D_{w}}{\partial p_{j}}$ if $g_{i}=g_{j}$). Its gradient ratio is
\begin{equation}
\begin{aligned}
\left|\frac{\partial D_{w}}{\partial p_{f}} \left/ \frac{\partial D_{w}}{\partial p_{b}} \right. \right| &= \frac{w_{f}\sum_{i}w_{i}(p_{i}+g_{i})-w_{f}\sum_{i}w_{i}p_{i}g_{i}}{w_{b}\sum_{i}w_{i}p_{i}g_{i}} \\
&=\frac{w_{f}}{w_{b}}(\frac{2}{1-D_{w}}-1) \geq \frac{w_{f}}{w_{b}},
\end{aligned}
\end{equation}
where $w_{f}$ and $w_{b}$ are the weights for foreground and background respectively. From Eq. (\ref{weighted_dice}), we can find three properties of weighted Dice loss. First, its gradient ratio can be adaptively changed due to the item $\frac{2}{1-D_{w}}-1$. Second, this change is amplified by the factor $\frac{w_{f}}{w_{b}}$. Third, the gradient ratio has a lower bound of $\frac{w_{f}}{w_{b}}$. These properties make it effective to improve the segmentation by tuning the weights $w_{f}$ and $w_{b}$.

The same properties can be found in Tversky loss,
\begin{equation}
T = 1 - \frac{\sum_{i=1}^{N}p_{i}g_{i}}{\sum_{i=1}^{N}p_{i}g_{i}+\alpha\sum_{i=1}^{N}p_{i}\left( 1-g_{i}\right) +\beta\sum_{i=1}^{N}\left( 1-p_{i}\right)g_{i}},
\end{equation} 
where $\alpha + \beta = 1$, so it can be rewritten as 
\begin{equation}
T = 1 - \frac{\sum_{i=1}^{N}p_{i}g_{i}}{\sum_{i=1}^{N}(\alpha p_{i} + \beta g_{i})}.
\end{equation}   
Its gradient can be calculated by
\begin{equation}
\frac{\partial T}{\partial p_{j}} = -\frac{g_{j}\sum_{i=1}^{N}(\alpha p_{i} + \beta g_{i})-\alpha \sum_{i}p_{i}g_{i}}{[\sum_{i=1}^{N}(\alpha p_{i} + \beta g_{i})]^{2}},
\end{equation}
and the gradient ratio is
\begin{equation}
\begin{aligned}
\left|\frac{\partial T}{\partial p_{f}} \left/ \frac{\partial T}{\partial p_{b}} \right. \right| &= \frac{\sum_{i=1}^{N}(\alpha p_{i} + \beta g_{i})-\alpha \sum_{i}p_{i}g_{i}}{\alpha \sum_{i}p_{i}g_{i}} \\
&=\frac{1}{\alpha}\frac{1}{1-T}-1 \geq \frac{\beta}{\alpha}.
\end{aligned}
\end{equation}
By tuning the importance of false positives and false negatives in the denominator, Tversky loss also can achieve a high sensitivity for peripheral small airways.

Despite the merits of these losses, they do not address the intra-class imbalance problem. The cost for a high tree length detected is the significant dilation problem, because the same large gradients are assigned to all foreground points. Actually, we do not need to increase the gradients to the easy-to-segment parts such as the large airways, while more attention should be paid to the hard-to-segmentation areas to improve the connectivity of the predicted bronchi. To this end, we propose a Root Tversky loss which combines the voxel-level focal function into Tversky loss, and a General Union loss is further generated by adding the distance-based weights.  

Root Tversky loss achieves the element-wise focal function by changing the root of the predictions in the numerator of Tversky loss,
\begin{equation}
T_{r} = 1 - \frac{\sum_{i=1}^{N}p_{i}^{r}g_{i}}{\alpha \sum_{i}p_{i}+ \beta \sum_{i}g_{i}},
\end{equation}   
where $0 < r < 1$ and $\alpha + \beta = 1$. It is differentiated in term of $p_{j}$, and the gradient is
\begin{equation}
\frac{\partial T_{r}}{\partial p_{j}} = - \frac{rg_{j}p_{j}^{r-1}\left( \alpha \sum_{i}p_{i}+ \beta \sum_{i}g_{i}\right) -\alpha \sum_{i}p_{i}^{r}g_{i}}{\left(\alpha \sum_{i}p_{i}+ \beta \sum_{i}g_{i} \right) ^{2}}.
\end{equation}
As $r-1<0$, we replace $p_{i}^{r-1}$ by $(p_{i}+\epsilon_{i})^{r-1}$ where $\epsilon$ is a small positive number. The same gradients are given to all background voxels while $p_{j}^{r-1}$ controls the gradients given to foreground locations. Since $0<r<1$, much larger gradient is given to the foreground voxel with small $p_{j}$. Its gradient ratio is
\begin{equation}
\begin{aligned}
\left|\frac{\partial T_{r}}{\partial p_{f}} \left/ \frac{\partial T_{r}}{\partial p_{b}} \right. \right| &= \frac{rp_{f,j}^{r-1} (\alpha \sum_{i}p_{i}+ \beta \sum_{i}g_{i}) - \alpha \sum_{i}p_{i}^{r}g_{i}}{\alpha \sum_{i}p_{i}^{r}g_{i}} \\ &= \frac{rp_{f,j}^{r-1}}{\alpha} \frac{1}{1-T_{r}}-1 \geq \frac{r}{\alpha}-1,
\end{aligned}
\end{equation}
where $p_{f,j}$ is the prediction of a foreground voxel j. Comparing Eq. (\ref{tversky}) and (\ref{r_tversky}), the main difference between Tversky loss and Root Tversky loss is that the amplification factor $\frac{1}{\alpha}$ in Tversky loss is replaced by $\frac{rp_{f,j}^{r-1}}{\alpha}$. In this new factor, the $p_{f,j}^{r-1}$ item achieves the voxel-wise focal function for the hard-to-segment points, while for the easy-to-segmentation airway voxels, this factor is near $\frac{r}{\alpha}$, resulting in smaller gradient ratio and alleviating the dilation problem.

As the intra-class imbalance is caused by the large number of thick airway voxels, we can manually decrease the weights for most large airway points. In this paper, we achieve this by assigning the weights for each voxel according to their distance to the centerlines,
\begin{equation}
w_{i} = \begin{cases}
1-m\left(\frac{d_{i}}{d_{max}} \right)^{r_{d}}, &  g_{i} = 1 \\
1, & g_{i} = 0 
\end{cases}
\end{equation}
where $d_{i}$ is the shortest distance from the current location to the centerline, $d_{max}$ is the maximum $d_{i}$ in one case and $r_{d}$ controls the pattern of decay. By adding this weight to Eq. (\ref{r_tversky}), the loss function is in the following form:
\begin{equation}
U = 1 - \frac{\sum_{i=1}^{N}w_{i}p_{i}^{r_{l}}g_{i}}{\sum_{i=1}^{N}w_{i} (\alpha p_{i}+ \beta  g_{i}) }.
\end{equation} 
To reduce the number of hyper-parameters, we set $m = \frac{1-\alpha}{\alpha}$ in $w_{i}$. Its gradient with regard to $p_{j}$ is
\begin{equation}
\frac{\partial U}{\partial p_{j}} = -\frac{w_{j}rg_{j}p_{j}^{r-1}\sum_{i}w_{i}\left( \alpha p_{i}+ \beta g_{i}\right) -w_{j}\alpha \sum_{i}w_{i}p_{i}^{r}g_{i}}{[\sum_{i=1}w_{i}(\alpha p_{i}+ \beta g_{i})]^{2}}.
\end{equation}
And the gradient ratio of this loss is
\begin{equation}
\begin{aligned}
\left|\frac{\partial U}{\partial p_{f}} \left/ \frac{\partial U}{\partial p_{b}} \right. \right| &= \frac{w_{f,j}rp_{f,j}^{r-1}\sum_{i}w_{i}\left( \alpha p_{i}+ \beta g_{i}\right) -w_{f,j}\alpha \sum_{i}w_{i}p_{i}^{r}g_{i}}{\alpha \sum_{i}w_{i}p_{i}^{r}g_{i}} \\ &= w_{f,j}\left[ \frac{rp_{f,j}^{r-1}}{\alpha} \frac{1}{1-U}-1 \right]\\ &\geq (1-m)(\frac{r}{\alpha}-1).
\end{aligned}
\end{equation}
It is seen that the distance-based weight $w_{i}$ is directly applied on the gradient ratio of each point. Besides, since the weights of most large airway voxels are decreased, small airways play a more important role in the item $\frac{1}{1-U}$. In other words, the adaptive changing of gradient ratio depends more on the sensitivity of peripheral bronchi. By adopting this distance-based weight, the dilation issue around the large airways can be resolved while keeping a high tree length detected.
\subsection{Segmentation Results}
\begin{figure*}[ht]	
	\centering
	\includegraphics[scale=0.45]{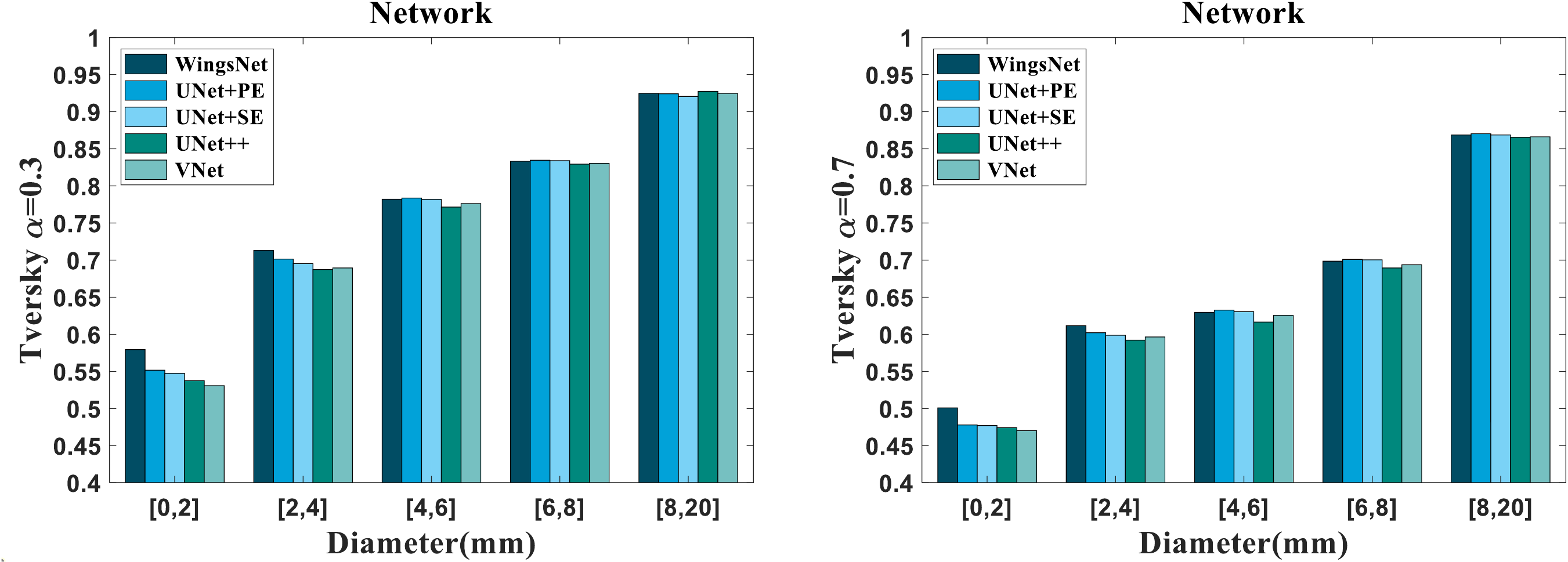}
	\caption{The segmentation results of five networks within different airway diameter intervals. Tversky index is used as the evaluation metric. The hyper-parameter $\alpha$ is $0.3$ in the left figure and $0.7$ in the right one.}
	\label{Network_Ts}
\end{figure*}
\begin{figure*}[ht]	
	\centering
	\includegraphics[scale=0.45]{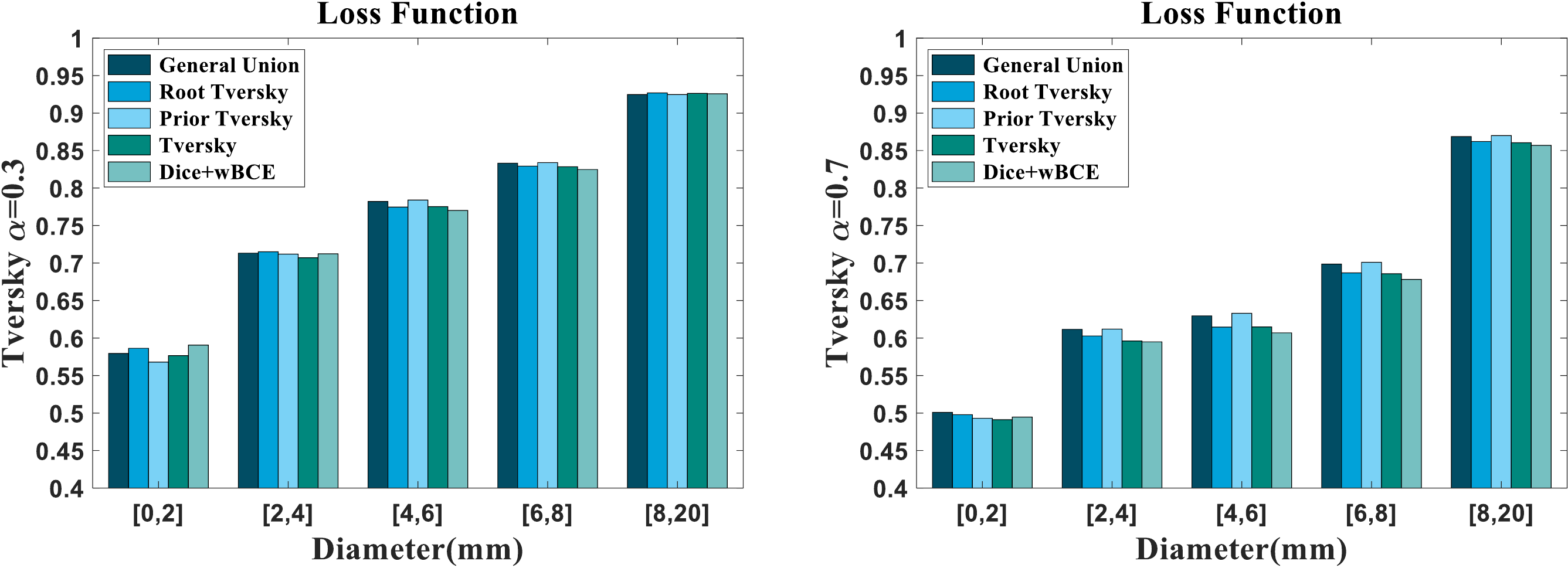}
	\caption{The segmentation results of five loss functions within different airway diameter intervals. Tversky index is used as the evaluation metric. The hyper-parameter $\alpha$ is $0.3$ in the left figure and $0.7$ in the right one.}
	\label{Loss_Ts}
\end{figure*}
We compare the segmentation performance of different network architectures by Tversky index. In this evaluation metric, when $\alpha<0.5$, more importance is given to the recall than precision, and vice versa. As shown in Fig.~\ref{Network_Ts}, in both situations ($\alpha=0.3$ and $\alpha=0.7$), the proposed WingsNet outperforms other network architectures within the diameter interval between 0 and 4mm. We also evaluate five loss functions by this metric. The results are illustrated in Fig.~\ref{Loss_Ts}. When $\alpha$ is $0.3$, Root Tversky loss achieves higher Tversky than General Union loss within the interval between 0 and 4mm. In contrast, when $\alpha=0.7$, General Union loss outperforms Root Tversky loss in all intervals. This means that GUL achieves higher precision than RTL while the latter can detect more peripheral branches. Prior Tversky loss achieves comparable performance with GUL except for the interval between 0 and 2mm, demonstrating the effectiveness of element-wise focal function for distal small airway segmentation. The combination of Dice loss and weighted Cross Entropy loss can obtain similar sensitivity for peripheral bronchi with GUL by tuning the weight in wBCE loss, but the dilation also becomes more serious in other branches, leading to the decrease in precision. 
\subsection{Hyper-parameters}
Actually, our method includes some hyper-parameters, and the selection of these parameters is a time-consuming project. To provide a clear illustration of their impacts, in this section, we set experiments on the Binary Airway Segmentation dataset to analyze each of them. We demonstrate the performance in two training stages. In the first stage, the the network is forced to achieve a high tree length detected, while in the second stage, the model is fine-tuned to increase the precision of segmentation. In this following experiments, the network was trained by General Union loss in the first stage, while Root Tversky loss was used in the second stage.
\subsubsection{$r_{l}$ in GUL/RTL}
\begin{figure}[ht]	
	\centering
	\includegraphics[scale=0.36]{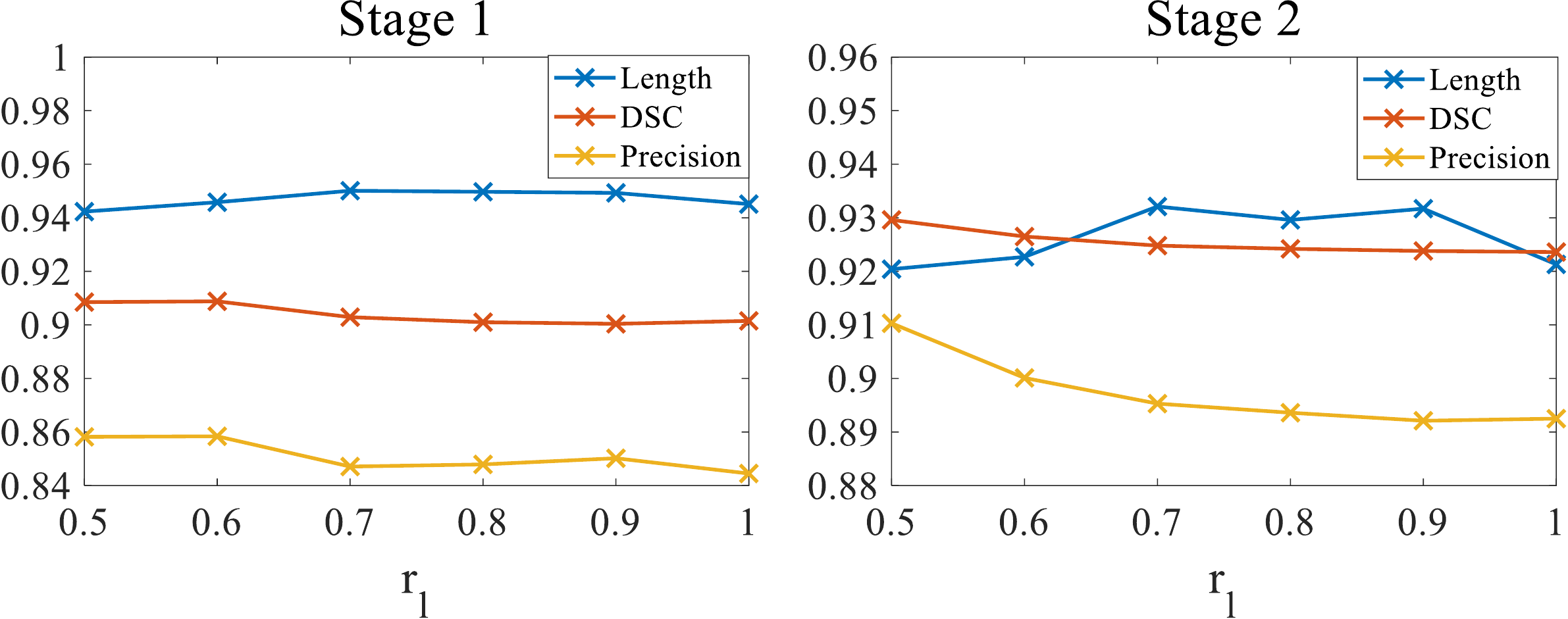}
	\caption{$r_{l}$ is the root of the prediction in the numerator of GUL/RTL, which controls the extent of the focal function. It also impacts the gradient ratio and a small $r_{l}$ can alleviate the dilation issue.}
	\label{rl}
\end{figure}
The hyper-parameter $r_{l}$ in GUL or RTL is the root of the predictions in the numerator, which controls the extent of the element-wise focal function. Loss function with smaller $r_{l}$ focuses more on the failure locations. Moreover, smaller $r_{l}$ leads to a smaller amplification factor of the gradient ratio. Therefore, as shown in Fig. \ref{rl}, the precision drops as $r_{l}$ rises. Besides, the excessive focus on the failure points is not beneficial to detect more small airways. In both stages, the length detected of the results of $r_{l}=0.5$ is slightly worse than the results of $r_{l}=1.0$. In contrast, a moderate focal effect helps the network learn better representations for peripheral bronchi. 
\subsubsection{$r_{d}$ in GUL}  
\begin{figure*}[ht]	
	\centering
	\includegraphics[scale=0.45]{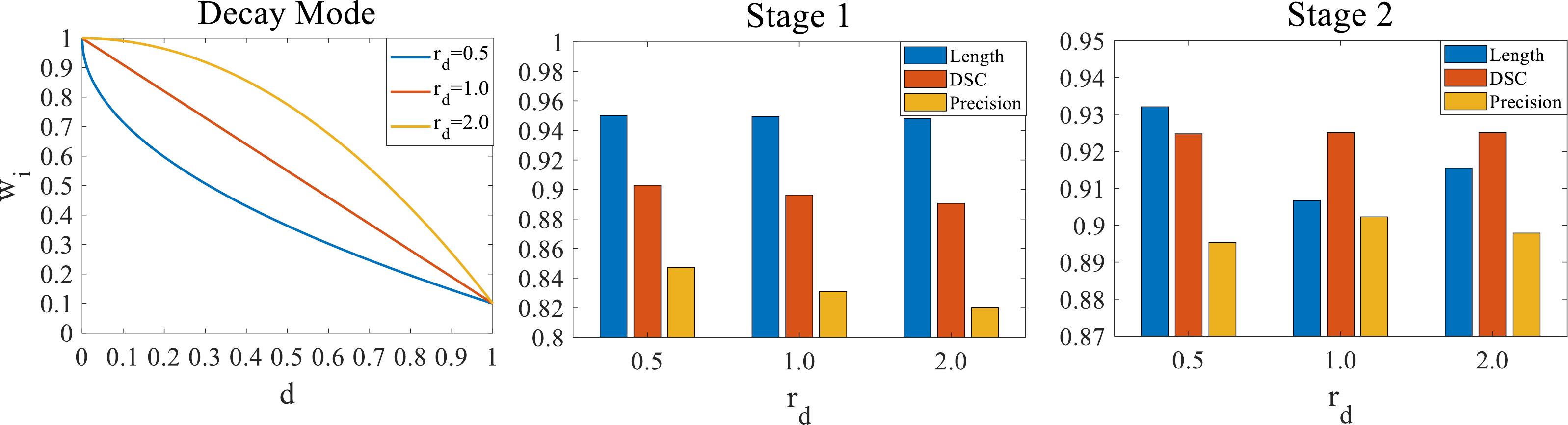}
	\caption{$r_{d}$ is the root of the relative distance in GUL, which decides the speed of the distance-based weight decay of airway voxels.}
	\label{rd}
\end{figure*}
The hyper-parameter $r_{d}$ in GUL controls the speed of the distance-based weight decay of airway voxels. In this experiment, we evaluated three decay modes as shown in Fig. \ref{rd}, $r_{d}=0.5$, $r_{d}=1.0$ and $r_{d}=2.0$. When $r_{d}=0.5$, the weight decreases fast around the centerline. A linear decay is achieved by $r_{d}=1.0$ and $r_{d}=2.0$ can keep a high weight around the centerline. In the first stage, all the modes can yield a high length detected while $r_{d}=0.5$ performs better than others for the gradient dilation problem. In the second stage, paying more attention to the centerlines can improve the length detected. Therefore, we chose $r=0.5$ in other experiments.
\subsubsection{$\alpha_{e}$ and $\alpha_{d}$}
\begin{figure}[ht]	
	\centering
	\includegraphics[scale=0.4]{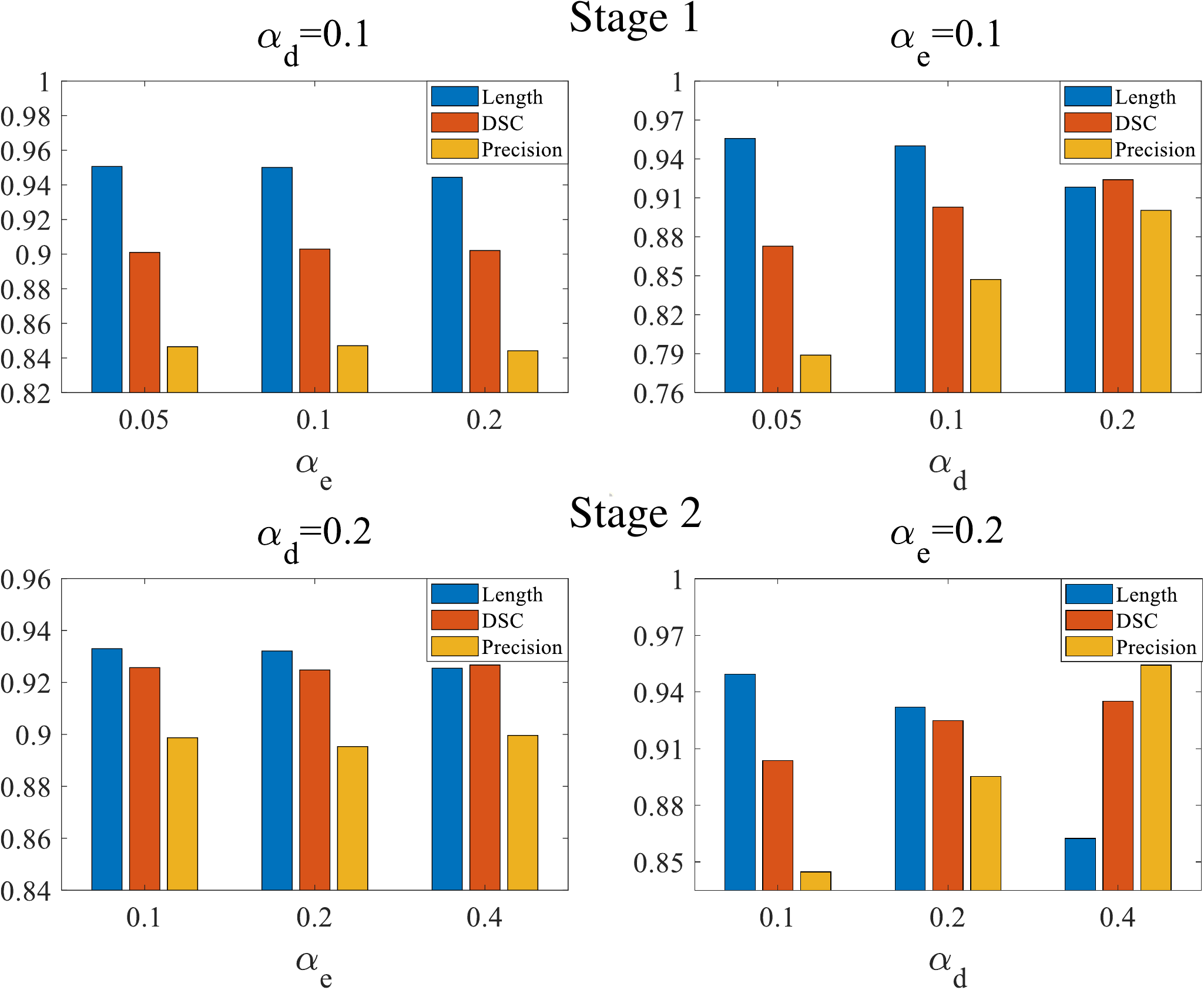}
	\caption{$\alpha_{e}$ and $\alpha_{d}$ are the hyper-parameters in GUL/RTL imposed on the encoding group and decoding group respectively. The final segmentation performance is mainly affected by $\alpha_{d}$. Small $\alpha$ leads to high gradient ratio, improving the length detected while decreasing the precision.}
	\label{ll_lr}
\end{figure}
In WingsNet, two loss functions are imposed on the encoding group and decoding group respectively. To evaluate different combinations of the hyper-parameters $\alpha_{e}$ and $\alpha_{d}$ in these two loss functions, we fix one parameter while varying another one in both stages. As shown in Fig. \ref{ll_lr}, since we use the predictions of the decoding group as the final results, the trade-off between sensitivity and specificity is mainly controlled by $\alpha_{d}$. Smaller $\alpha_{d}$ results in a higher amplification factor of the gradient ratio, alleviating the gradient erosion and improving the length detected. In contrast, the impact of $\alpha_{e}$ is insignificant in the first stage, while in the second stage, a large $\alpha_{e}$ reduces the length detected. In test set, the combination of $\alpha_{e}=0.1$ and $\alpha_{d}=0.2$ achieves the best results in the second stage. However, in validation set, the optimal selection is $\alpha_{e}=\alpha_{d}=0.2$. Thus, in other experiments, we chose $\alpha_{e}=\alpha_{d}=0.1$ in the first stage and $\alpha_{e}=\alpha_{d}=0.2$ in the second stage.
\subsubsection{$p_{d}$ in SpatialDropout}
\begin{figure}[ht]	
	\centering
	\includegraphics[scale=0.36]{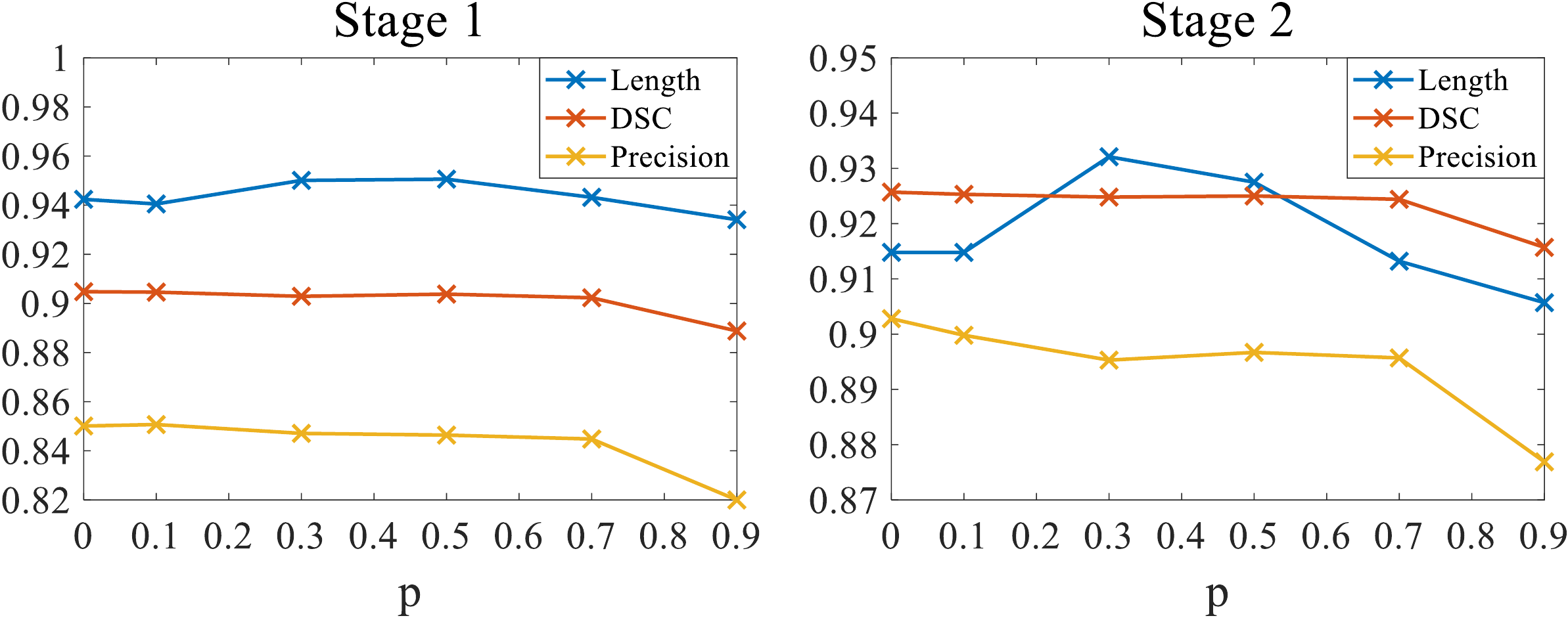}
	\caption{$p_{d}$ in SpatialDropout controls the probability of each channel to be dropped. Small $p_{d}$ can improve the generalization while large $p_{d}$ affects the learning of hierarchical representations.}
	\label{p_sd}
\end{figure}
When performing SpatialDropout, a hyper-parameter $p_{d}$ is set to control the probability of each channel to be dropped. As shown in Fig. \ref{p_sd}, a suitable $p_{d}$ helps the network to learn robust features for the distal small airways, improving the length detected. However, when $p_{d}$ is greater than 0.5, both the length detected and precision drop as $p_{d}$ increases. In this case, the prediction only relies on several layers, which is a little like the deeply supervised nets, limiting the learning of hierarchical representations. When choosing this hyper-parameter, $p_{s}=0.3$ achieves the best length detected in the second stage in both the validation set and test set. Besides, our method performs stably within the interval of $[0.3,0.5]$. Thus, we selected $p_{d}=0.3$ in other experiments.
\subsubsection{Threshold in Data Augmentation}
\begin{figure}[ht]	
	\centering
	\includegraphics[scale=0.38]{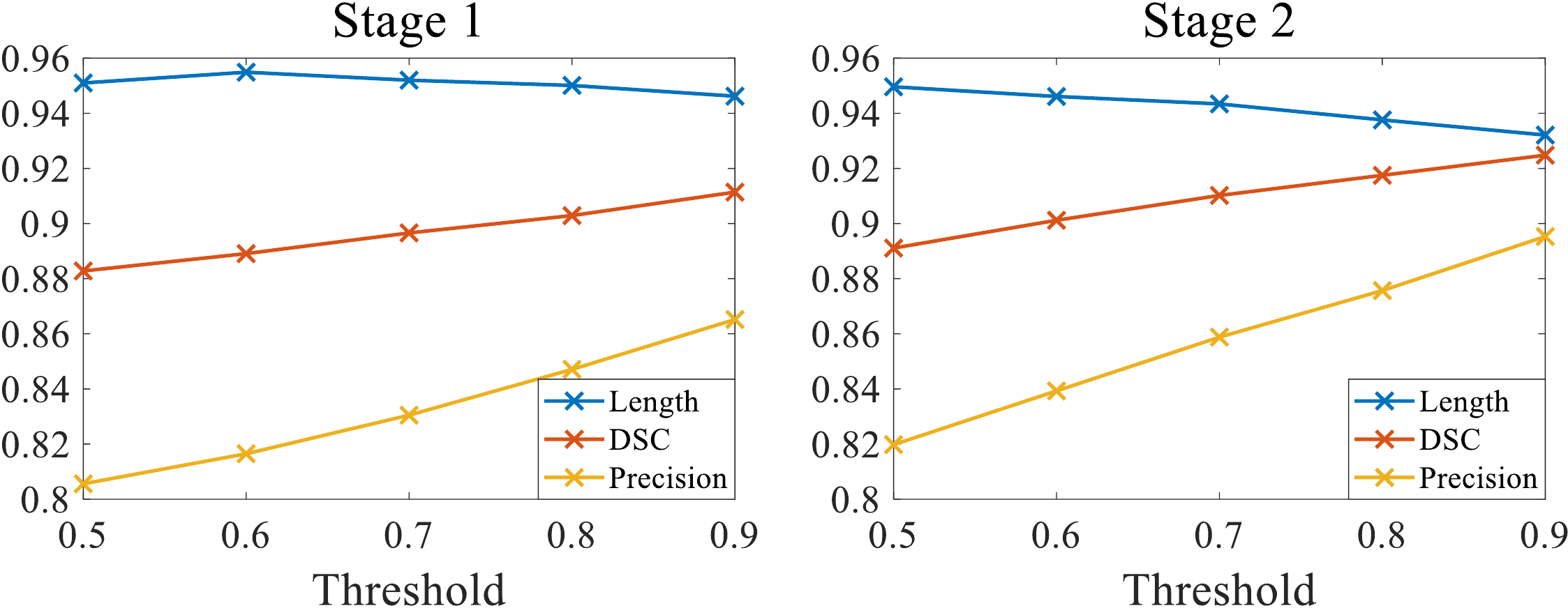}
	\caption{Threshold in data augmentation. This parameter affects the thickness of small airway annotations after random rotation. A large threshold slightly decreases the length detected while improving the precision.}
	\label{threshold}
\end{figure}
In data augmentation, the threshold of interpolation directly affects the thickness of small airway annotations after random rotation. A small threshold can enlarge the bronchi annotations, which is beneficial to detect more small airways but leads to more serious dilation problem. As shown in Fig. \ref{threshold}, the opposed trends are seen in the lines of length detected and precision. To improve the precision of our results, we chose a pretty high threshold ($0.9$) in the second stage while slightly decreasing this number to $0.8$ for higher sensitivity in the first stage. 
\subsubsection{$p_{s}$ in Data Sampling}
\begin{figure}[ht]	
	\centering
	\includegraphics[scale=0.38]{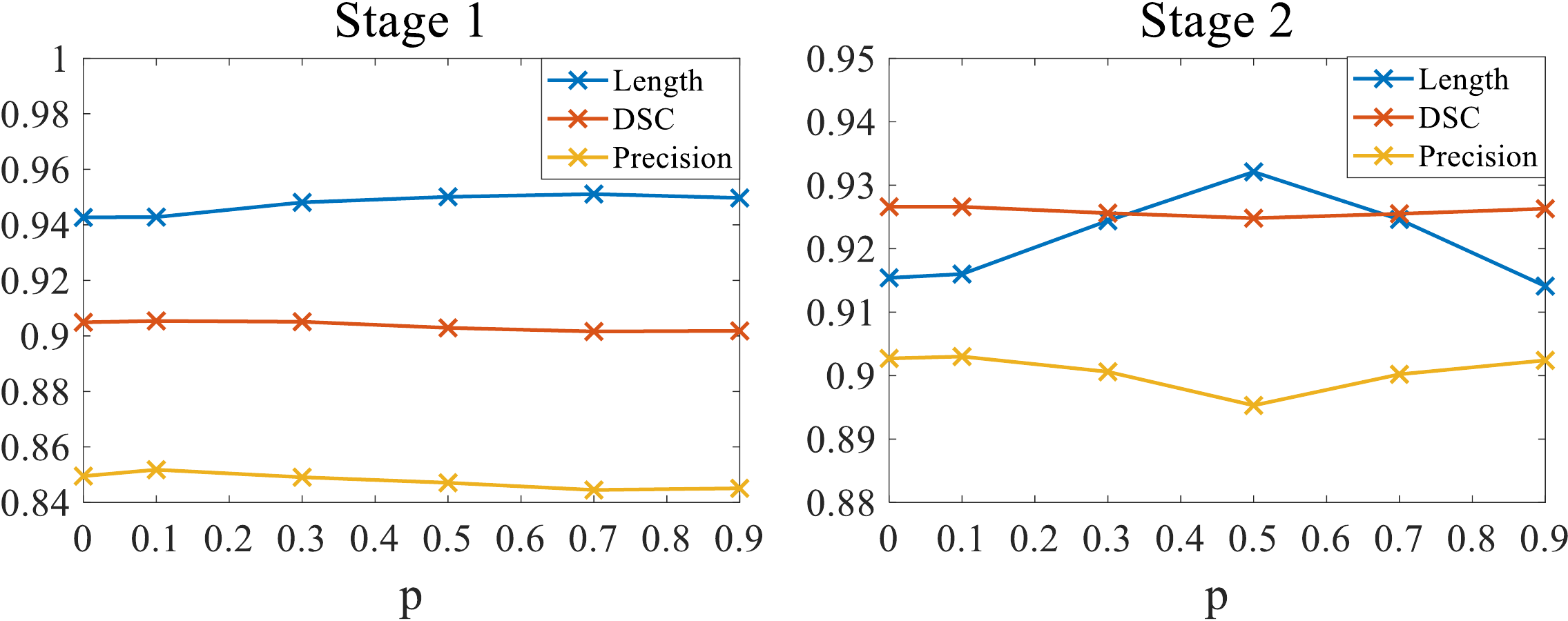}
	\caption{During training, a probability $p_{s}$ is set to perform hard skeleton sampling or random sampling. Adopting hard skeleton sampling can boost the length detected while the overuse inversely affects the performance.}
	\label{p_hss}
\end{figure}
During training, a probability $p_{s}$ is set to perform hard skeleton sampling or random sampling. As shown in Fig. \ref{p_hss}, in the first stage, adopting hard skeleton sampling can improve the length detected about $1\%$, and this margin is more significant (about $2\%$) in the second stage. It is also seen that a pretty high $p_{s}$ leads to the decrease in both length detected and precision. In this case, the model over-fits to these difficult regions, affecting the generalization capability. Therefore, we selected $p_{s}=0.5$ in other experiments.

\end{document}